\documentclass{aa}
\usepackage{graphicx}
\usepackage{txfonts}
\usepackage{natbib}
\usepackage{hyperref} 
\usepackage{dsfont}
\hypersetup{colorlinks=true,linkcolor=blue,citecolor=blue,urlcolor=blue}
\usepackage{orcidlink}
\usepackage{lscape}

\newcommand{\gaia}{{\it Gaia }}
\begin{document}






\title{ (Re)mind the  gap:  a hiatus in  star formation history \\ unveiled by APOGEE DR17   }

\author { E. Spitoni \orcidlink{0000-0001-9715-5727}\inst{1}  \thanks {email to: emanuele.spitoni@inaf.it}, F. Matteucci \orcidlink{0000-0001-7067-2302} \inst{1,2,3}    \and R. Gratton \orcidlink{0000-0003-2195-6805}  \inst{4}
\and B. Ratcliffe \orcidlink{0000-0003-1124-7378}\inst{5}\and I. Minchev \orcidlink{0000-0002-5627-0355}\inst{5} \and G. Cescutti \orcidlink{0000-0002-3184-9918} \inst{1,2,3}   }
\institute{I.N.A.F. Osservatorio Astronomico di Trieste, via G.B. Tiepolo
 11, 34131, Trieste, Italy  
   \and Dipartimento di Fisica, Sezione di Astronomia,
  Universit\`a di Trieste, Via G.~B. Tiepolo 11, I-34143 Trieste,
  Italy
 \and  I.N.F.N. Sezione di Trieste, via Valerio 2, 34134 Trieste, Italy
\and I.N.A.F.  Osservatorio Astronomico di Padova, Vicolo dell’Osservatorio 5, 35122 Padova, Italy
\and Leibniz-Institut f\"ur
   Astrophysik Potsdam, An der
  Sternwarte 16, 14482, Potsdam, Germany 
}

 \date{Received xxxx / Accepted xxxx}

\abstract {The analysis of several spectroscopic surveys indicates the presence of a bimodality  between the disc stars in the abundance ratio space of [$\alpha$/Fe] versus [Fe/H]. The two stellar groups are commonly referred to as the high-$\alpha$ and low-$\alpha$ sequences.  Some models capable of reproducing such a bimodality, invoke the presence of a hiatus in the star formation history in our Galaxy, whereas other models explain the two sequences by means of stellar migration.}
{Our aim is to show that the existence of the gap in the star formation rate between high-$\alpha$ and low-$\alpha$ is evident in the stars of APOGEE DR17, if one plots [Fe/$\alpha$] versus [$\alpha$/H], thus confirming previous suggestions by \citet{gratton1996}  and  \citet{fuhr1998}. Then we try to interpret the data by means of detailed chemical models. 
}{We compare the APOGEE DR17  red giant stars with the predictions of a detailed chemical evolution model based on the two-infall paradigm, taking also into account possible accretion of dwarf satellites.}
{The APOGEE DR17 abundance ratios [Fe/$\alpha$] versus [$\alpha$/H]  exhibit a sharp increase of [Fe/$\alpha$] at a nearly constant [$\alpha$/H]  (where $\alpha$ elements considered are Mg, Si, O) during the transition between the two disc phases. This observation strongly supports the hypothesis that a hiatus in star formation occurred during this evolutionary phase. Notably, the most pronounced growth in the [Fe/$\alpha$] versus [$\alpha$/H] relation is observed for oxygen, as this element is exclusively synthesised in core-collapse supernovae. The revised version of the two-infall chemical evolution model proposed in this study reproduces the APOGEE DR17 abundance ratios  better than before. Particularly noteworthy is the model's ability to predict the hiatus in the star formation between the two infalls of gas,  which form the thick and thin disc, respectively, thus generating abundance ratios compatible with APOGEE DR17 data.}{We show that the signature of a hiatus in the star formation is imprinted in the APOGEE DR17 abundance ratios. A chemical model predicting a stop in the star formation of a duration of roughly 3.5 Gyr, and where the high-$\alpha$ disc starts  forming from  pre-enriched gas by a previous encounter with a dwarf galaxy can well explain the observations.}


\titlerunning{(Re)mind the gap}

\authorrunning{Spitoni et al.}

\maketitle

\section{Introduction}

The analysis of ground-based spectroscopic surveys, including the Galactic Archaeology with HERMES survey (GALAH; \citep{buder2021}), the Apache Point Observatory Galactic Evolution Experiment project  \citep[APOGEE,][]{nidever2015,Majewski:2017ip,Ahumada2019, apogeedr172022,imig2023}, the Gaia-ESO project (e.g., \citealt{RecioBlanco:2014dd,kordopatis2015,RojasArriagada:2016eq,RojasArriagada:2017ka}), the AMBRE project \citep{Mikolaitis:2017gd,delaverny2013}, indicates the presence of   two distinct sequences of disc stars in the abundance ratio space of [$\alpha$/Fe] versus [Fe/H]. These sequences are commonly referred to as the high-$\alpha$ and low-$\alpha$. Recent confirmation of this separation comes from space-based \gaia DR3 mission \citep{recioDR32022b}  with the  General Stellar Parametrizer - spectroscopy (GSP-Spec, \citealt{recioDR32022a}) and with \gaia XP spectra \citep{chandra2023}.

In our comprehension of the Milky Way (MW)'s evolutionary journey, our Galaxy does not exist in isolation, undergoing significant events of gas accretion from the intergalactic medium.
Various theoretical models describing the evolution of Galactic discs propose a tight connection between bimodality and a delayed accretion episode involving pristine or metal-poor gas.  \citet{spitoni2019, spitoni2020, spitoni2022} demonstrated that a substantial delay of approximately 4 Gyr between two consecutive gas accretion episodes is necessary to account for the dichotomy observed in the local APOKASC sample (APOGEE+ Kepler Asteroseismology Science Consortium, \citealt{pinso2014}) within the solar neighbourhood \citep{victor2018} and APOGEE DR16 stars.

Specifically, their predictions indicate that the star formation rate (SFR) reaches a minimum at an age of around 8 billion years since the beginning of star formation. A comparable suppression of star formation  at the age of 8 Gyr was also identified by \citet{snaith2015}, utilizing chemical abundances from \citet{adi2012} and isochrone ages from \citet{haywood2013} for solar-type stars. \citet{katz2021} concluded that APOGEE data exhibits a dilution of the interstellar medium from 6 kpc to the outskirts of the disc, occurring prior to the beginning of low-$\alpha$ formation.
In the context of the two-infall model and the shock-heating theory, \citet{noguchi2018} also proposed a significant time gap between the two accretion phases. According to their model, an initial infall event leads to the formation of the high-$\alpha$ sequence, succeeded by a pause until the shock-heated gas in the Galactic dark matter halo undergoes radiative cooling and becomes available for accretion by the Galaxy. Within this framework, \citet{noguchi2018} determined that the Star Formation Rate (SFR) of the Galactic disc exhibits two distinct peaks with an approximate separation of 5 billion years.

Through the analysis of ESO/HARPS spectra from local solar twin stars, \citet{nissen2020} identified a distinct age-metallicity distribution with two populations, suggesting evidence for two gas accretion episodes onto the Galactic disc with an interceding quenching of star formation, simarly to  the scenario proposed by \citet{spitoni2019,spitoni2020} but taking into account also an enriched infalling gas. 
Also the recent AMBRE:HARPS data were reproduced by \citet{palla2022} using a chemical evolution model characterised by a two-infall scenario with a significant delay (3.25 Gyr) between the accretion episodes. 

Subgiant stars analysed by \citet{xiang2022} with the Large Sky Area Multi-Object Fibre Spectroscopic Telescope (LAMOST; \citealt{zheng2021}) confirmed a stellar age-metallicity distribution that bifurcates into two almost disjoint parts, separated at an age of approximately 8 Gyr. In \citet{sahlholdt2022}, the authors highlighted an age-metallicity relation characterised by several disconnected sequences, potentially associated with different star formation regimes throughout the MW disc evolution.

\citet{spitoni2022} demonstrated that the signature of a delayed gas infall episode, leading to a hiatus in the star formation history of the Galaxy, is evident both in the [Mg/Fe] versus [Fe/H] relation and in the vertical distribution of [Mg/Fe] abundances in the solar vicinity. The works of \citet{lian2020, lian2020b} presented an alternative version of the two-infall model, where a continuous episode of gas accretion is interrupted by two rapidly quenched starbursts. The first starburst contributes to the formation of the high-$\alpha$ thick disc, followed by the second.

In \citet{recio2024}, an analysis of the Kiel diagram for GSP Spec Gaia DR3 giant stars revealed a disc bimodality for the first time, eliminating concerns related to interstellar absorption. This bimodality gives rise to dual Red Giant Branch sequences and Red Clump characteristics in mono-metallicity populations, providing support for modeling the chemical evolution of the disc through distinct infall episodes.

 The presence  of a phase of (very) low stellar formation between the  high- and low-$\alpha$ sequences is also found in \citet{chandra2023}.
 In order to explain the orbital circularity and the metallicity of stars analysed with Gaia  by \citet{chandra2023}, a star formation  that peaked at early times and with a progressive exhaustion of the gas is required. This is consistent with what is shown by Figure 3 of \citet{xiang2022}. This shape of the star-forming history of the early disc explains some of low-populated regions in the [Fe/H]-circularity plane and resembles the hiatus in star formation proposed in the two-infalls model by \citet{chiappini1997} (much before Gaia data were available) and also by \citet{spitoni2019}.
 
 More recently,  \citet{dubay2024} confirmed that the star-formation history plays a crucial role in generating a bimodal [$\alpha$/Fe] distribution visible in APOGEE DR17 data. Testing the effects of  different delay time distributions (DTDs) for Type Ia SNe, they concluded that the DTD  alone cannot produce a bimodal [$\alpha$/Fe] distribution, confirming the findings of \citet{matteucci2009} and \citet{palla2021}. Finally, \citet{dubay2024}  claimed that  radial migration  does not reproduce the [$\alpha$/Fe] bimodality of APOGEE data, in agreement with   \citet{johnson2021} results, but in tension  with findings from \citet{sharma2021} and  \citet{chen2023}.

 The existence of a double sequence has been discussed in a cosmological framework in recent years. \citet{grand2018} clearly point out that a bimodal distribution in the [Fe/H]- [$\alpha$/Fe] plane may be a consequence of a significantly lowered gas accretion rate at ages between 6 and 9 Gyr. 
On the other hand,  \citet{buck2020}  suggested  that 
dichotomy in the  $\alpha$-sequence is a generic consequence of a gas-rich merger occurred at a certain epoch in the evolution of the Galaxy, which destabilised the gaseous disc at high redshift. Finally,  in the cosmological simulations presented by  \citet{agertz2021}, the authors concluded that the low-$\alpha$ sequence in outer Galactic regions (R> 6 kpc) has been assembled after a last major merger by the accretion of a metal-poor gas filaments  which is inclined with respect to the main Galactic plane.

As previously mentioned, the star formation history projected by the two-infall model by \citet{spitoni2019} predicts an extended phase with a hiatus.  \citet{gratton1996,gratton2000} and \citet{fuhr1998}  proposed that the abundance ratios [Fe/$\alpha$] versus [$\alpha$/H], where $\alpha$ = O, Mg, serve as a better diagnostic tool for revealing gaps in the star formation process using spectroscopic data.
In their examination of O and Fe abundances in a sample of stars within the solar neighbourhood, \citet{gratton1996,gratton2000} observed a rise in the [Fe/O] ratio by approximately 0.2 dex, while the [O/H] ratio remained constant during the transition from thick to thin disc phases. This suggested a sudden decline in star formation in the solar neighbourhood during that period. In fact,  elements synthesised exclusively by Core-Collapse Supernovae (CC-SNe), such as oxygen, should  stop  to be produced during the hiatus, whereas elements such as Fe, which is originating mainly from Type Ia SNe on long timescales, continue to be produced and ejected. As a consequence of that, we should observe an increase in [Fe/$\alpha$] at a fixed [$\alpha$/H]. 
These authors also proposed that if the increase in [Fe/O] while maintaining a constant [O/H] is linked to a sudden reduction in star formation during the transition from thick to thin disc phases, a corresponding void should be observable in the distribution of stars with [Fe/O]. \citet{fuhr1998} confirmed the existence of a gap in the SFR by analysing the [Fe/Mg] ratio in solar vicinity stars.

The primary objective of the present article is to re-examine the abundance ratios [Fe/$\alpha$] versus [$\alpha$/H], where $\alpha$ = O, Mg, and Si, in APOGEE DR17 stars. A comparative analysis with a revised version of the two-infall model presented by \citet{spitoni2019} will be conducted. We aim at discussing similarities and differences that emerge between data and model regarding the signatures of a cessation in star formation, as suggested by the abundance ratios.

While the existence of radial stellar migration is well-established \citep[see, e.g.,][]{roskar2008, schoenrich2009MNRAS, loebmn2011, minchev2012, kubryk2013}, its magnitude remains a subject of ongoing debate. Some works \citep[i.e.,][]{Nidever:2014fj,sharma2021} have suggested that the dichotomy observed in the [$\alpha$/Fe] ratios in thick and thin disc, can be explained through the stellar migration.   In their view, the thick disc is formed just by stars migrated from the innermost regions of the disc.   \citet{ratcliffe2023} proposed that high-$\alpha$ formed in inner disc before a steepening in [Fe/H] gradient (maybe due to a merger), and then migrated out and that is why we can see the high-$\alpha$ stars in the solar vicinity. It is worth noting that in the chemo-dynamical model of \citet{minchev2013}, which incorporates star formation history and chemical enrichment based on a classical one-infall chemical evolution scenario,    the radial migration cannot produce a gap in the [$\alpha$/Fe] vs. [Fe/H] abundance plane.  This has been confirmed by \citet{Khoperskov2020}, who suggested  the $\alpha$-dichotomy is strictly linked to different star formation regimes over the Galaxy evolution.

 To summarise,  many Galactic formation scenarios have been proposed to explain the disc bimodality in the chemical space. One hypothesis suggests that this dichotomy emerged due to a temporary halt in star formation \citep{spitoni2019, noguchi2018, Khoperskov2020}. Alternatively, efficient stellar migration could have redistributed stars, leading to the observed bimodality in the solar vicinity \citep{sharma2021, chen2023}. In this paper, we explore the first possibility in detail, bearing in mind that some of the high-$\alpha$ sequence stars observed today in the solar vicinity may have migrated from the innermost regions of the Galaxy.

Our paper is organised as follows: in Section \ref{apogee}, we present the APOGEE DR17 sample used in this work.
In Section \ref{models_sec}, the  adopted chemical evolution models are described.  In Section  \ref{results_sec},
 we present our results and finally, our conclusions  are drawn in Section \ref{conclu_sec}.

\begin{figure*}
\centering
\includegraphics[scale=0.24]{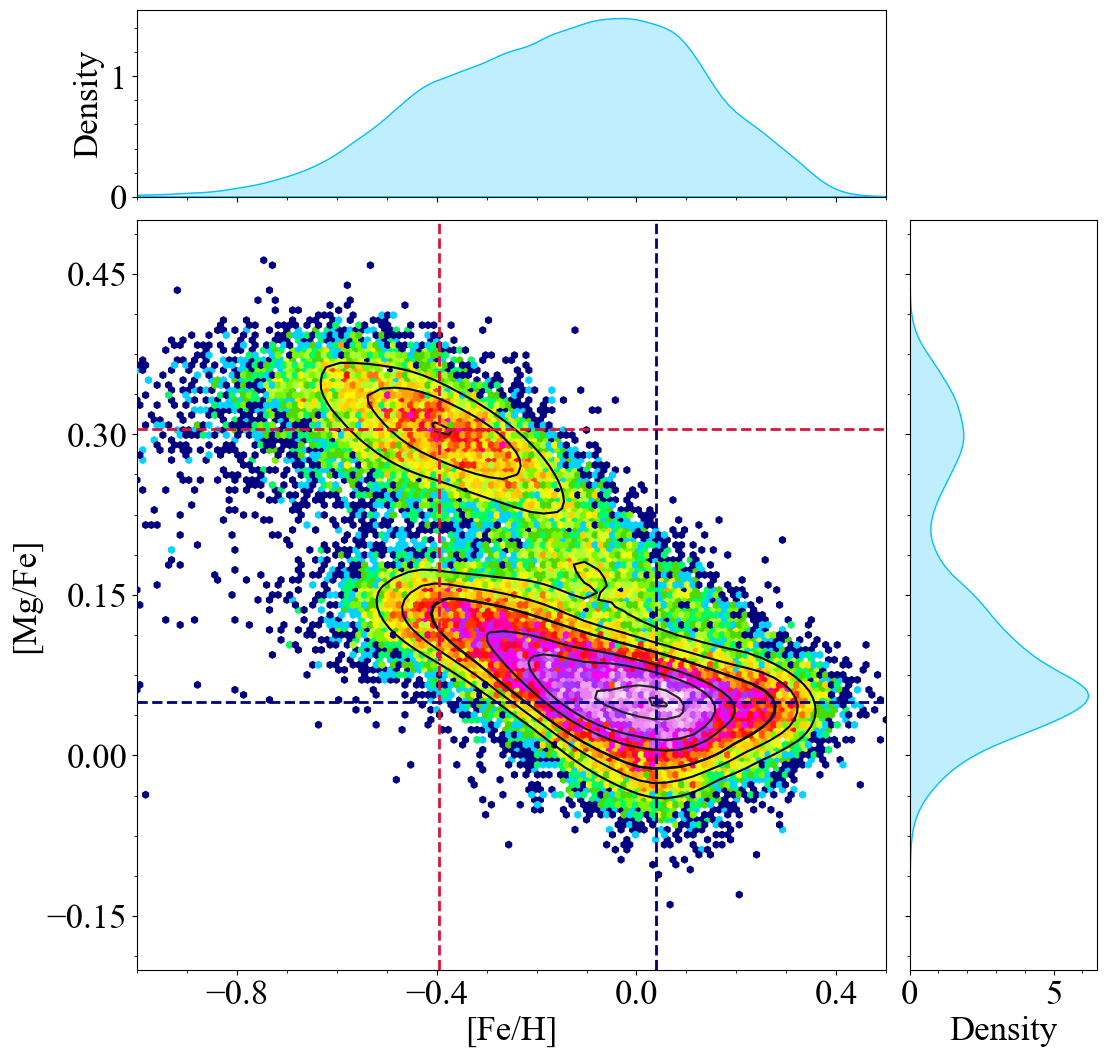}
\includegraphics[scale=0.24]{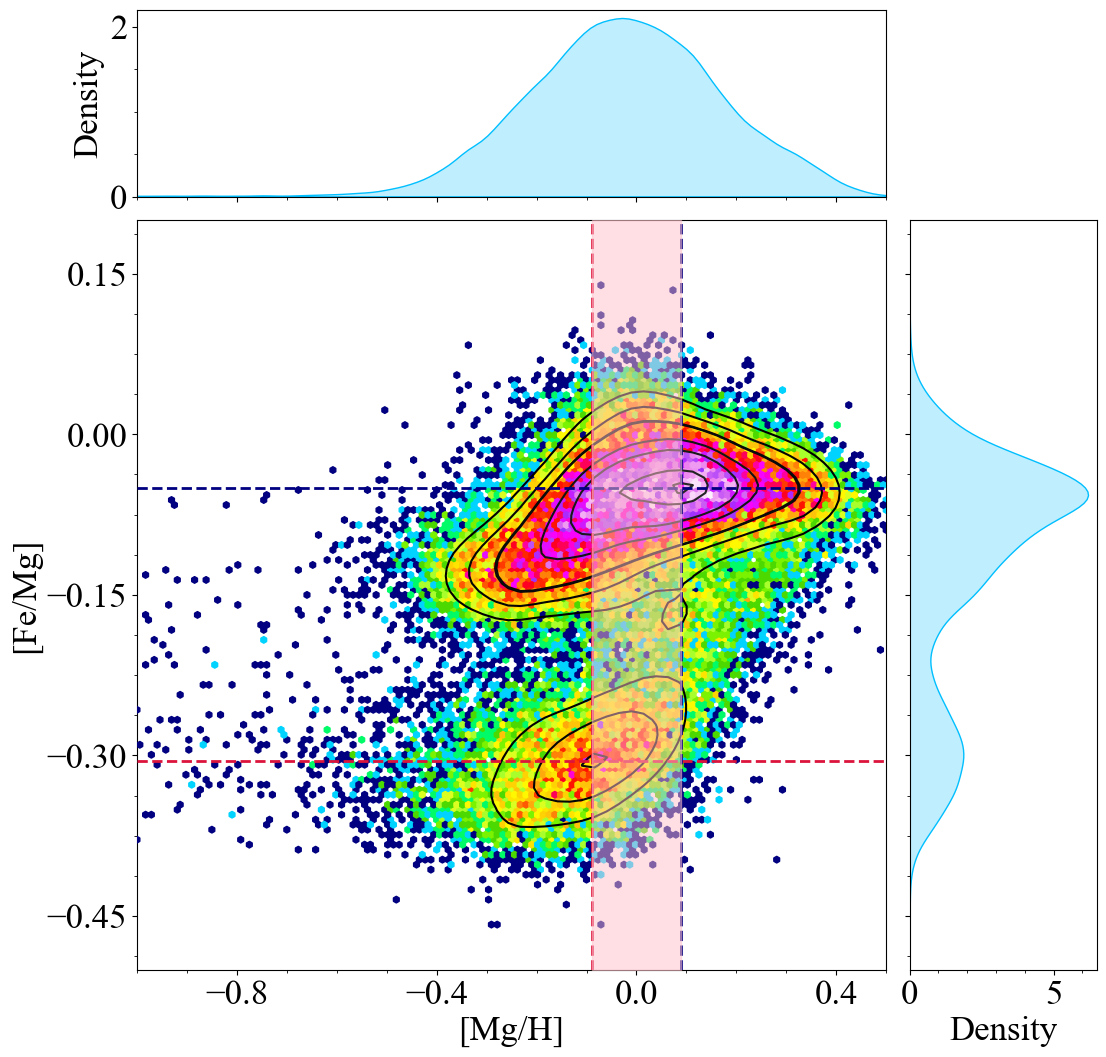}
\includegraphics[scale=0.24]{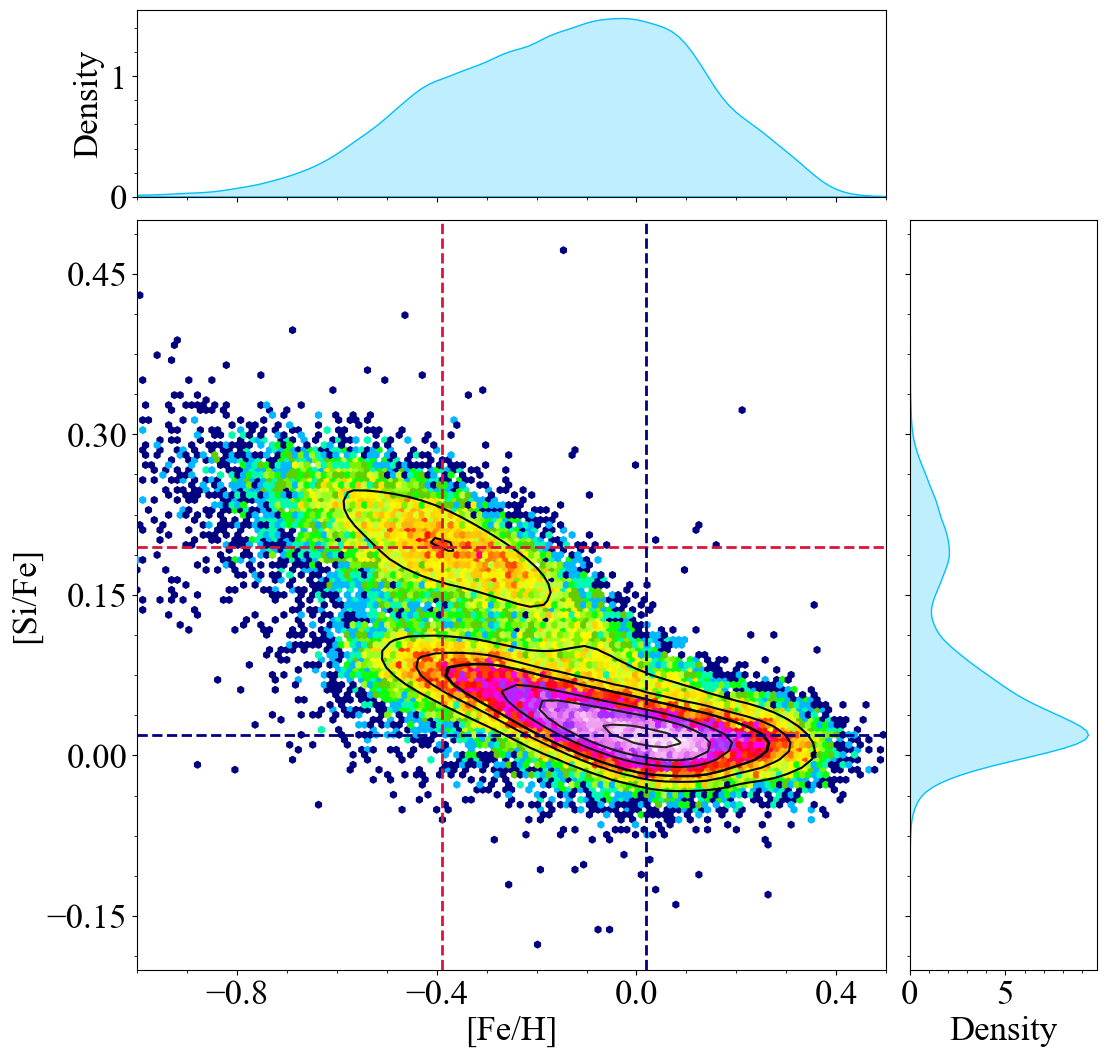}
\includegraphics[scale=0.24]{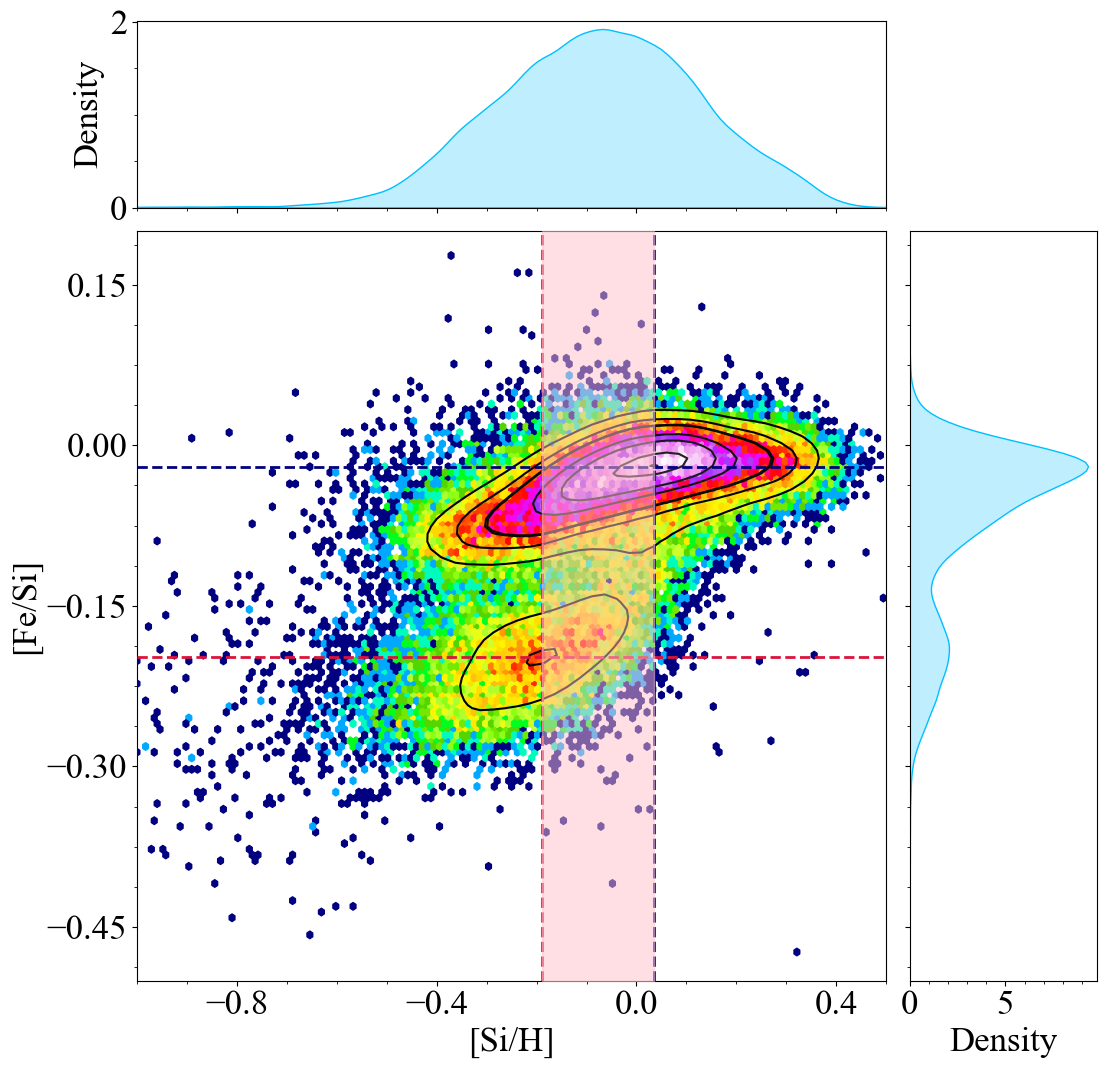}
\includegraphics[scale=0.24]{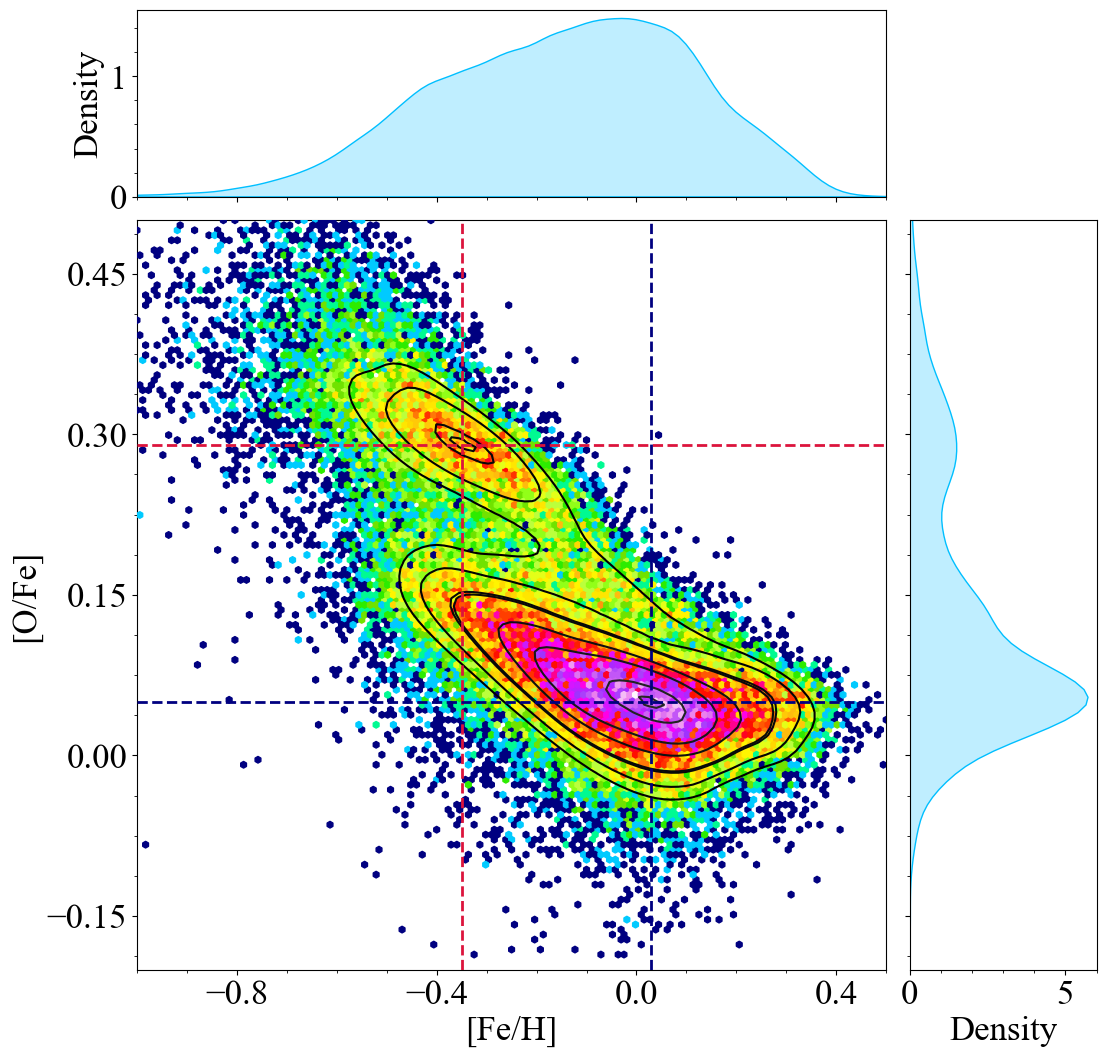}
\includegraphics[scale=0.24]{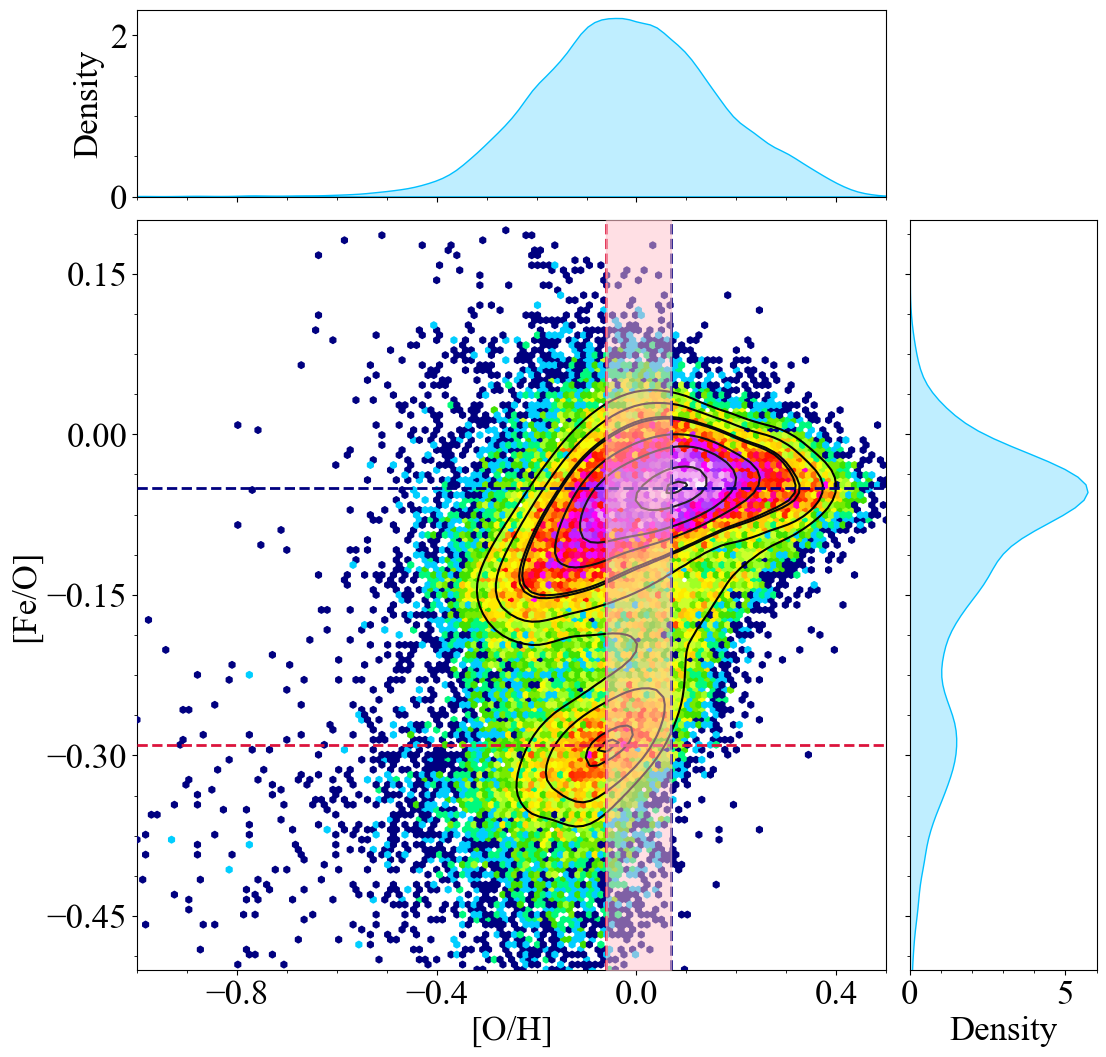}
  \caption{  Signatures of the hiatus in the star formation unveiled by APOGEE DR17 red giant stars.
  Density distribution of the observed [$\alpha$/Fe] versus [Fe/H] (left panels) and [Fe/$\alpha$] versus [$\alpha$/H] (right panels) abundance ratios  (see text for data selection details) where $\alpha$ = Mg (first row), $\alpha$ = Si (second row), and $\alpha$ = O (last row)  are reported. On the sides of each panel, the normalised Kernel Density Estimation (KDE) calculated with a Gaussian kernel of the abundance ratio distributions is also reported. The coordinates of the densest regions for the high-$\alpha$ and low-$\alpha$ stars are pinpointed with red and blue lines, respectively. In the right panels, the shaded pink area spans the region of the quantity $\Delta$[$\alpha$/H]$_{\text{peak}}$ as defined in eq. (\ref{eq:peak}).  The contour density lines correspond to the number of stars within bins in the specified abundance spaces, while the color coding is presented in logarithmic scale.
}
\label{data_density}
\end{figure*}

\section{The APOGEE DR17 data sample}
\label{apogee}
We utilise the high-resolution spectroscopic survey APOGEE DR17 \citep{apogeedr172022},
  APOGEE is part of the Sloan Digital Sky Surveys (SDSS) and uses the du Pont Telescope and the Sloan Foundation 2.5 m Telescope \citep{gunn2006} at Apache Point Observatory. 
Stellar parameters and abundances are determined using the APOGEE Stellar Parameters and Chemical Abundance Pipeline (ASPCAP; \citealt{gperez2016}). The primary objective of APOGEE is to create a comprehensive chemodynamical map  of all the structural components of the the Milky Way via near-twin, multiplexed NIR high-resolution spectrographs operating simultaneously in both hemispheres (APOGEE-N and APOGEE-S spectrographs respectively; both described in \citealt{wilson2019}). Extensive descriptions of the target selection and strategy are found  in \citet{zaso2017}. The model atmosphere used in APOGEE DR17 is MARCS atmospheres of \citet{gustafsson2008}   \citep[see][]{jonsson2020}. The line list is described in \citet{smith2021}.

 We considered stars with a signal-to-noise ratio $(S/N) >80$, a logarithm of surface gravity  $\log g<3.5$, Galactocentric distances enclosed between 7 and 9 kpc, and vertical heights $\lvert z \rvert<$ 2 kpc. 
The Galactocentric distances are determined according to \citet{leung2019}, available in the astroNN catalogue for APOGEE DR17 stars\footnote{\href{https://data.sdss.org/sas/dr17/apogee/vac/apogee-astronn}{https://data.sdss.org/sas/dr17/apogee/vac/apogee-astronn}}.

\begin{figure}
\centering
 \includegraphics[scale=0.39]{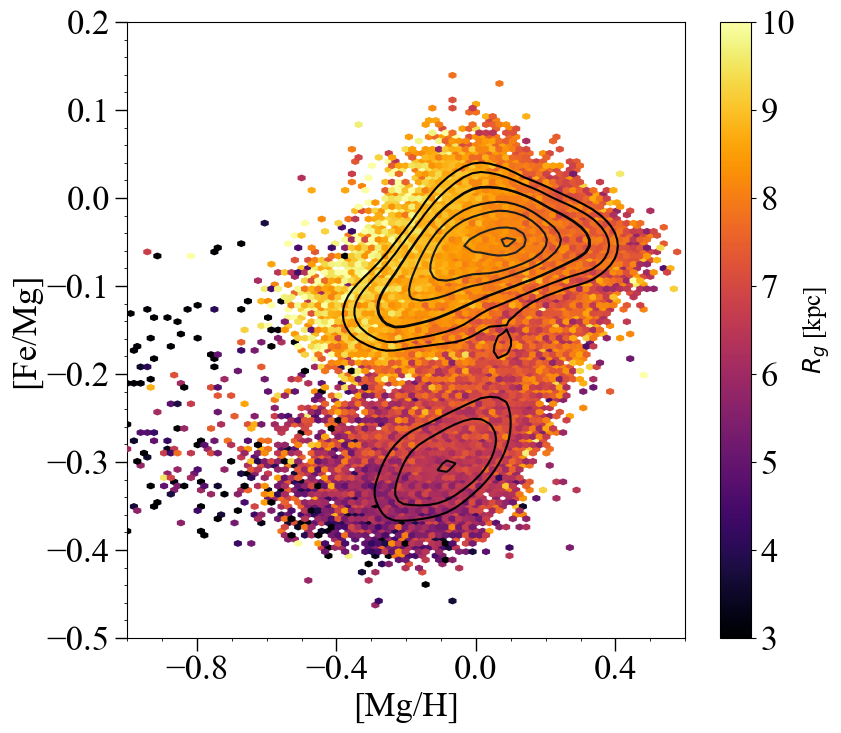}
 \caption{Observed stellar [Fe/Mg] versus [Mg/H] abundance ratios from APOGEE DR17 as in the upper right panel of Fig. \ref{data_density} colour coded with the median values of the guiding radii as computed by the value added astroNN catalogue. }
\label{guiding_radii}
\end{figure}

\begin{figure}
\centering
 \includegraphics[scale=0.3]{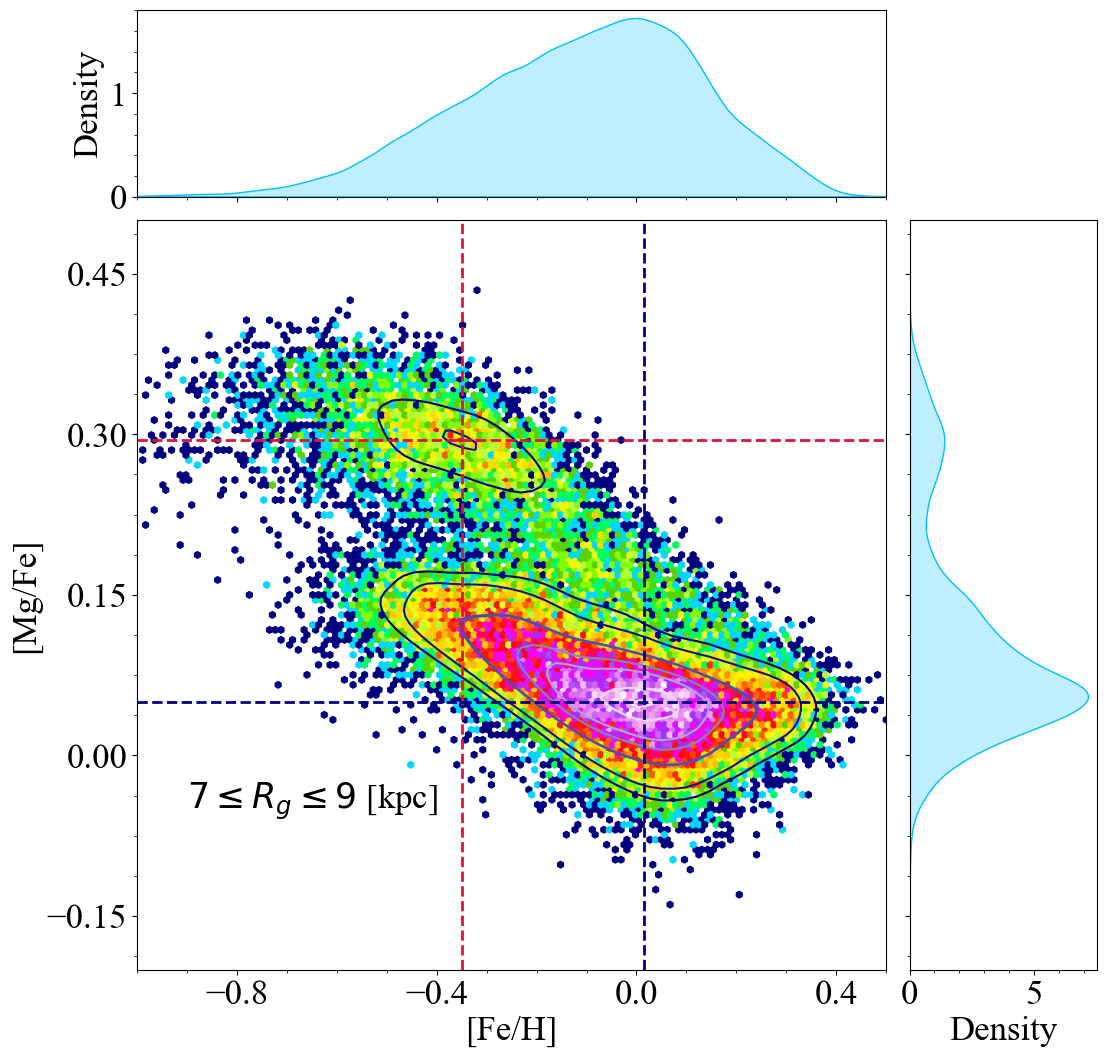}
 \includegraphics[scale=0.3]{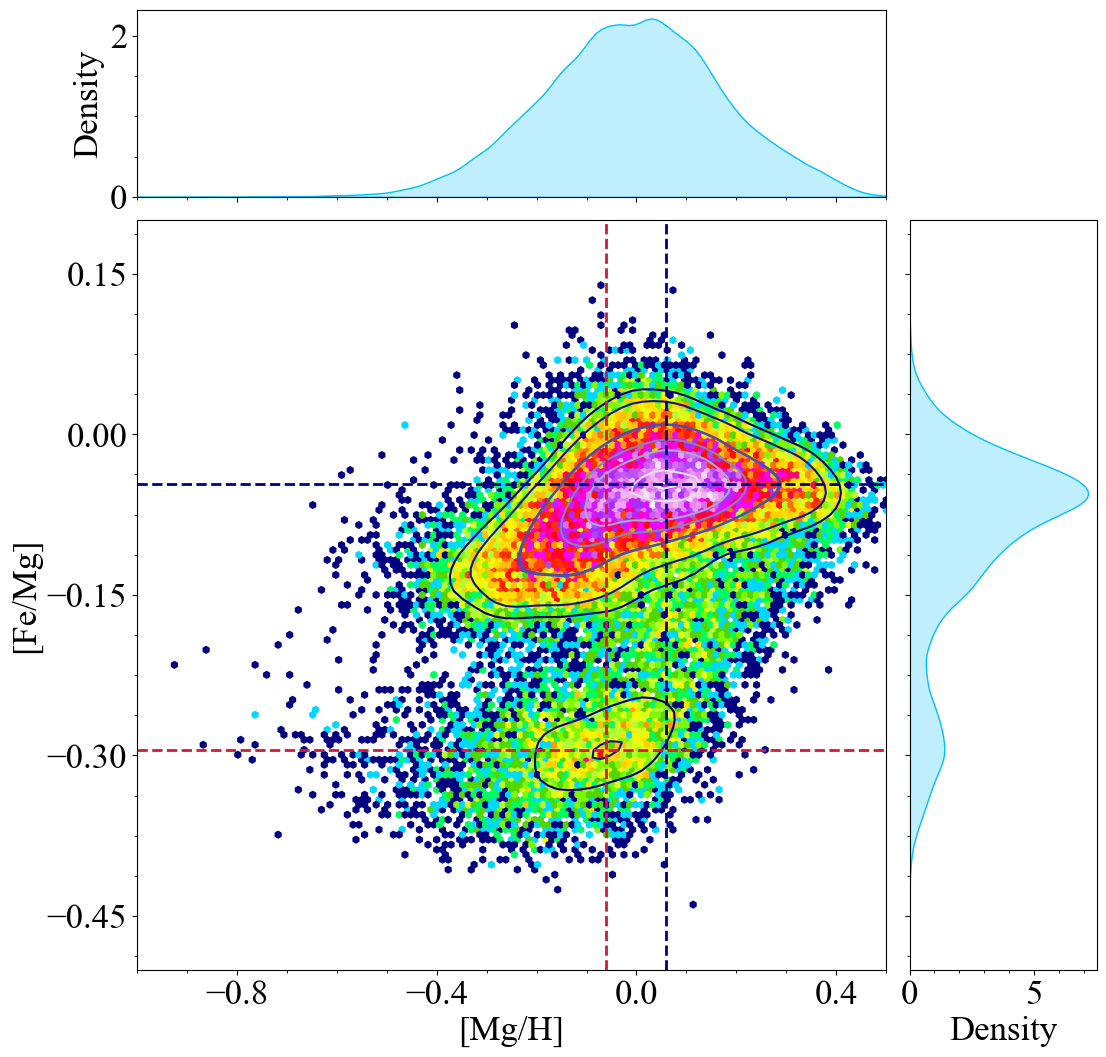}
  \caption{As the upper panels of  Fig. \ref{data_density} but for APOGEE DR17 stars with guiding radii $R_g$ between 7 and 9 kpc  (we do not impose condition on the Galactocentric distance as we did for Fig. \ref{data_density}) and vertical height $\lvert z \rvert<$ 2 kpc.}
\label{sel_gui}
\end{figure}

The accurate distances for distant stars were obtained using a deep neural network developed by \citet{leung2019}, trained with parallax measurements of nearby stars shared between $Gaia$ and APOGEE. This network enables the determination of spectro-photometric distances for APOGEE stars. The total number of considered stars is 47105. In Fig. \ref{data_density}, the observed [$\alpha$/Fe] versus [Fe/H] and [Fe/$\alpha$] versus [$\alpha$/H] abundance ratios from APOGEE DR17, where $\alpha$ = Mg, Si, and O, depict a clear bimodality.  Figure \ref{guiding_radii}  shows [Fe/Mg] versus [Mg/H], color-coded with guiding radii  as computed by the astroNN catalogue. In agreement with previous findings, a fraction of the stars on the high-$\alpha$ sequence are likely to have migrated from the inner Galactic regions, indicated by small guiding radii. In Fig. \ref{sel_gui}, the abundance ratio for stars with guiding radii enclosed between 7 and 9 kpc (without imposing any condition on
the Galactocentric distances) and vertical height $\lvert z \rvert<$ 2 kpc is presented, revealing a clear bimodality in this case as well.   The total number of stars  of the considered sample is 36484.  Although the number of stars in the high-$\alpha$ sequence is  smaller then in Fig. \ref{data_density}, the position of the two peaks (high-$\alpha$ and low-$\alpha$) does not change.

It is important to stress that in the present paper we will compare the above-mentioned abundance ratios with predictions of a model designed for the solar neighbourhood, where the thick and thin discs are formed sequentially. However, as highlighted by the results of the multi-zone chemical evolution model of \citet{spitoni2021} the chemical enrichment of the high-$\alpha$ is similar at different Galactocentric distances, and the results we will present will be still valid if stellar migration is included \citep{palla2022}.

\subsection{The SF hiatus  from observed abundance ratios   }
\label{peaks_obs}
In Fig. \ref{data_density}, the left panels depict the [$\alpha$/Fe] versus [Fe/H] abundance ratios (for $\alpha$ = Mg, Si, O), while the right panels illustrate [Fe/$\alpha$] versus [$\alpha$/H], showcasing the abundance ratios of APOGEE DR17 stars under the selection criteria outlined in the previous section. A distinct bimodality is evident for all considered elements across the panels.

\begin{figure*}
\centering
\includegraphics[scale=0.24]{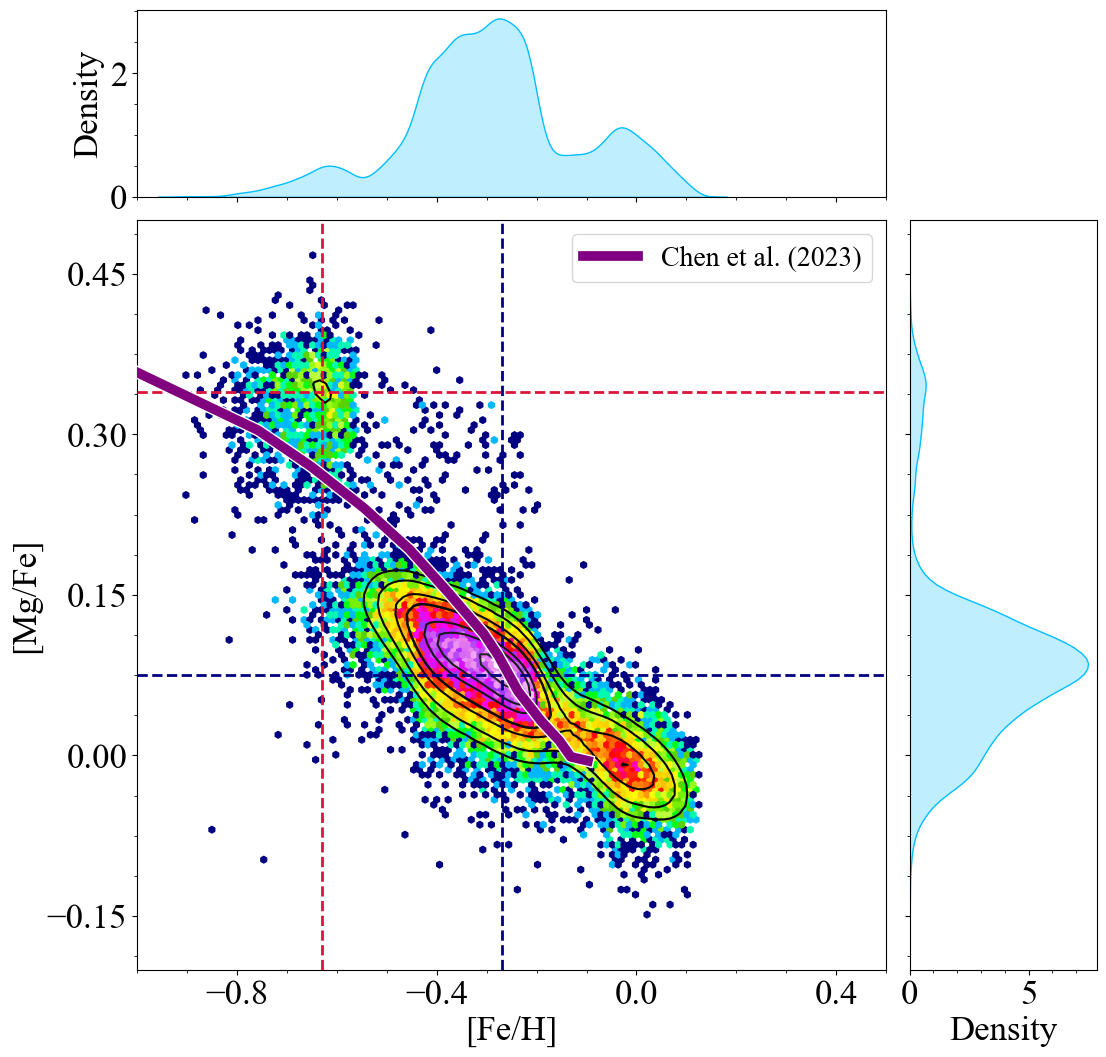}
\includegraphics[scale=0.24]{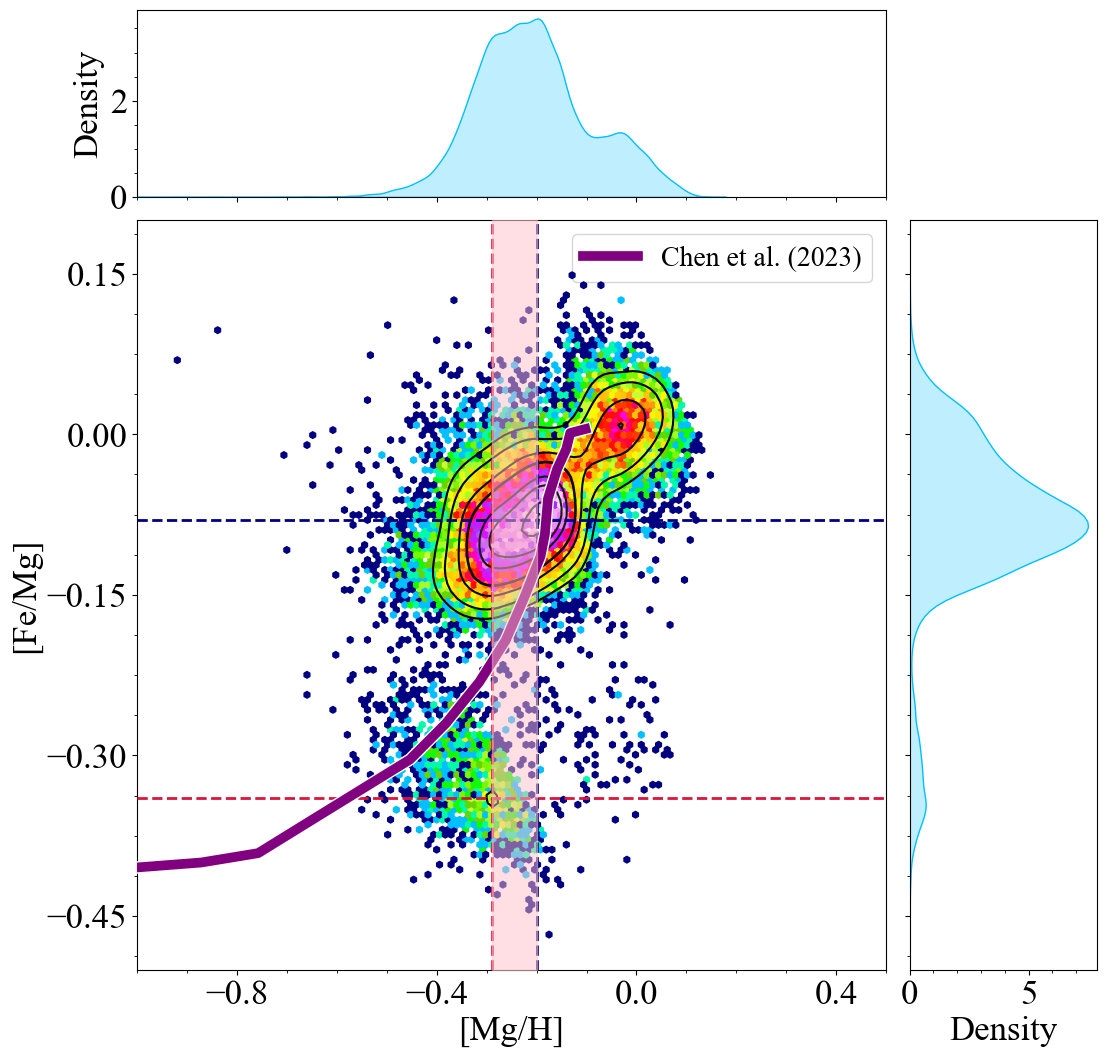}
\includegraphics[scale=0.24]{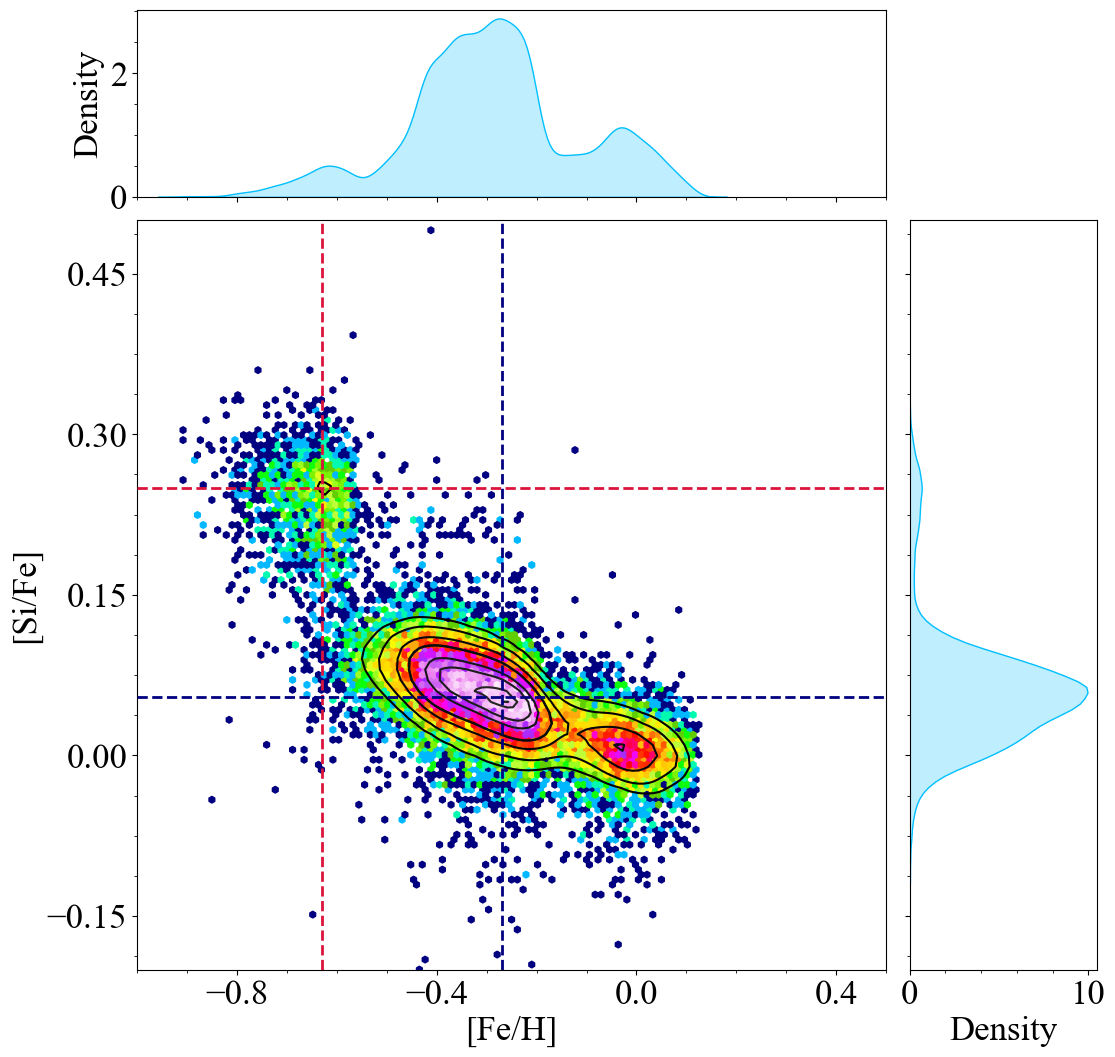}
\includegraphics[scale=0.24]{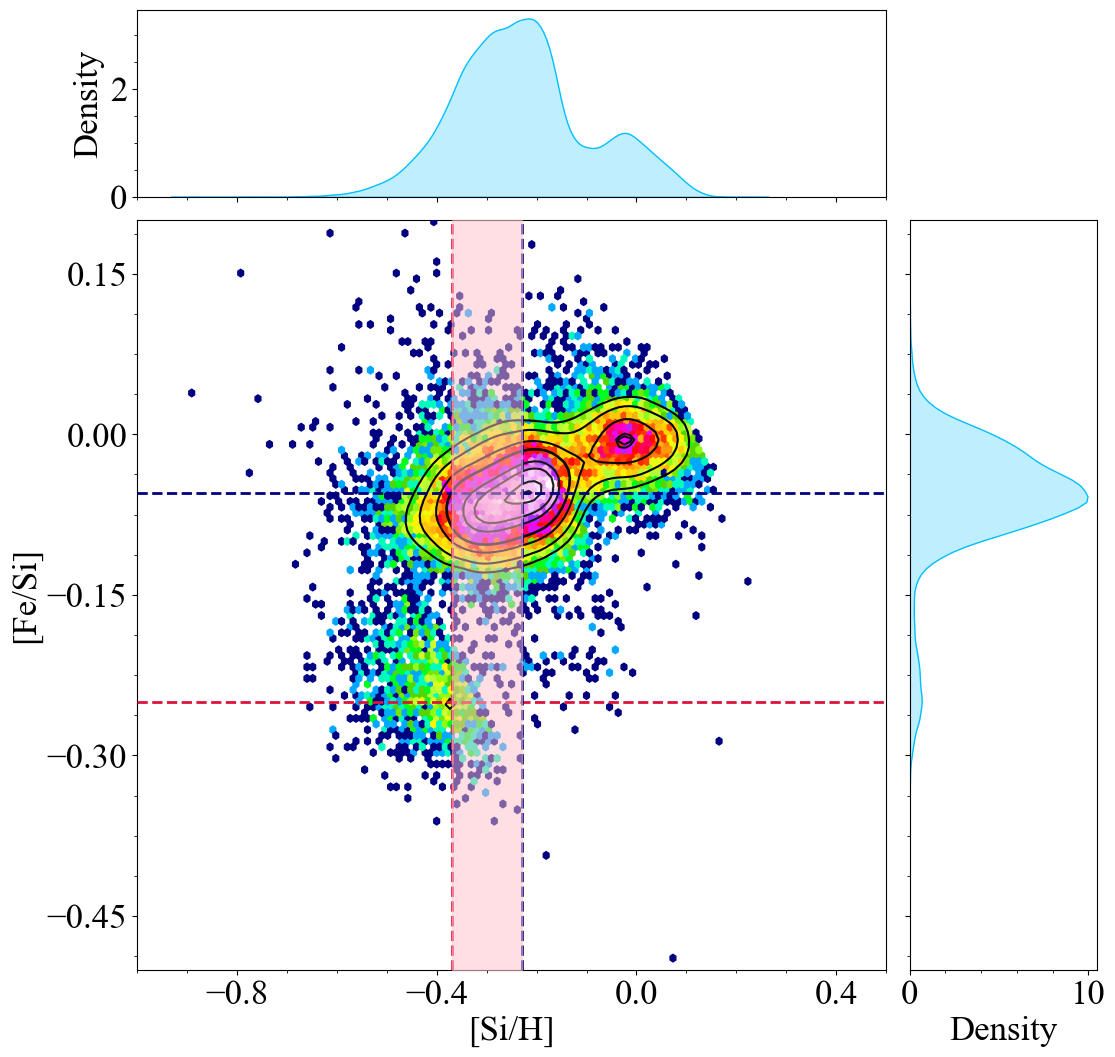}
\includegraphics[scale=0.24]{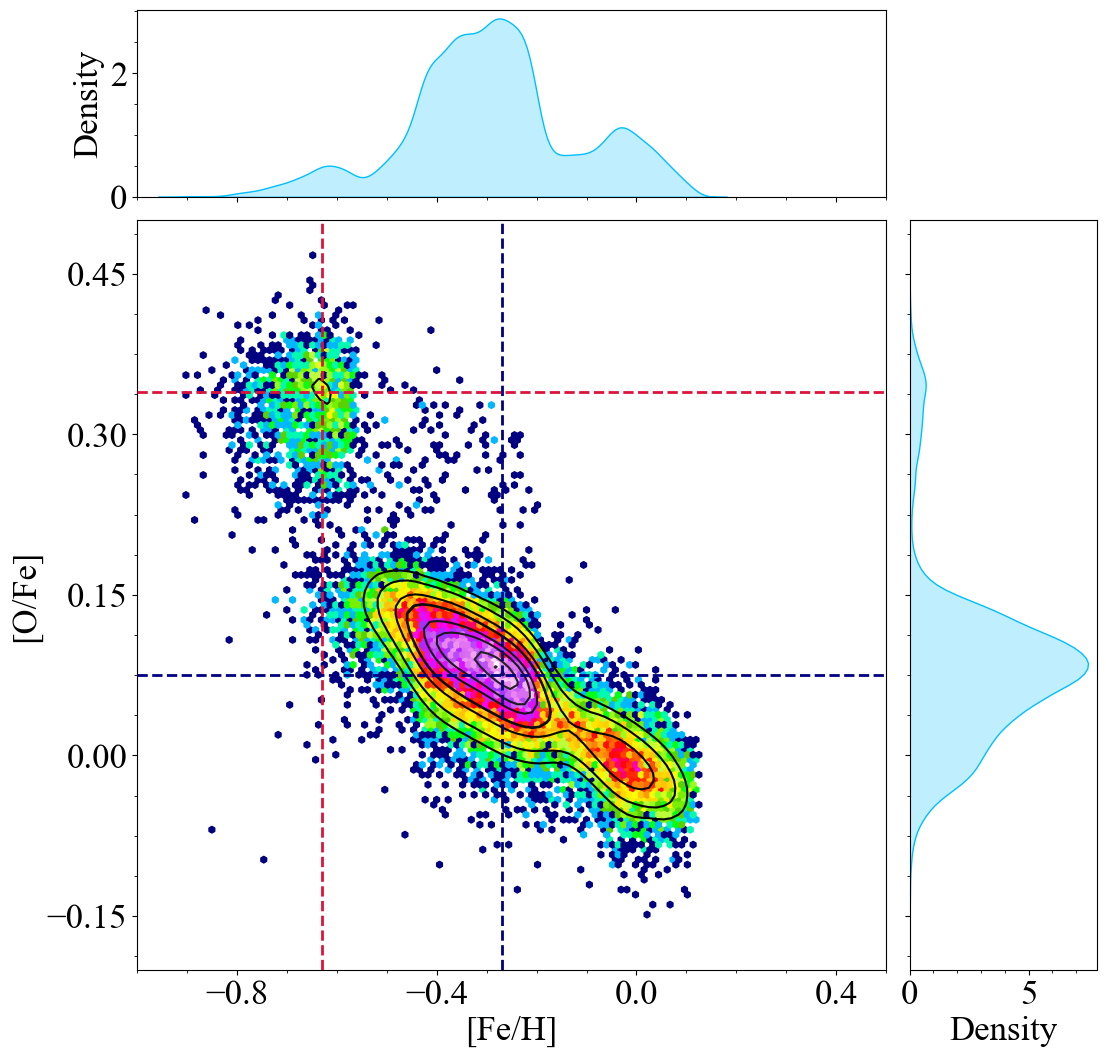}
\includegraphics[scale=0.23]{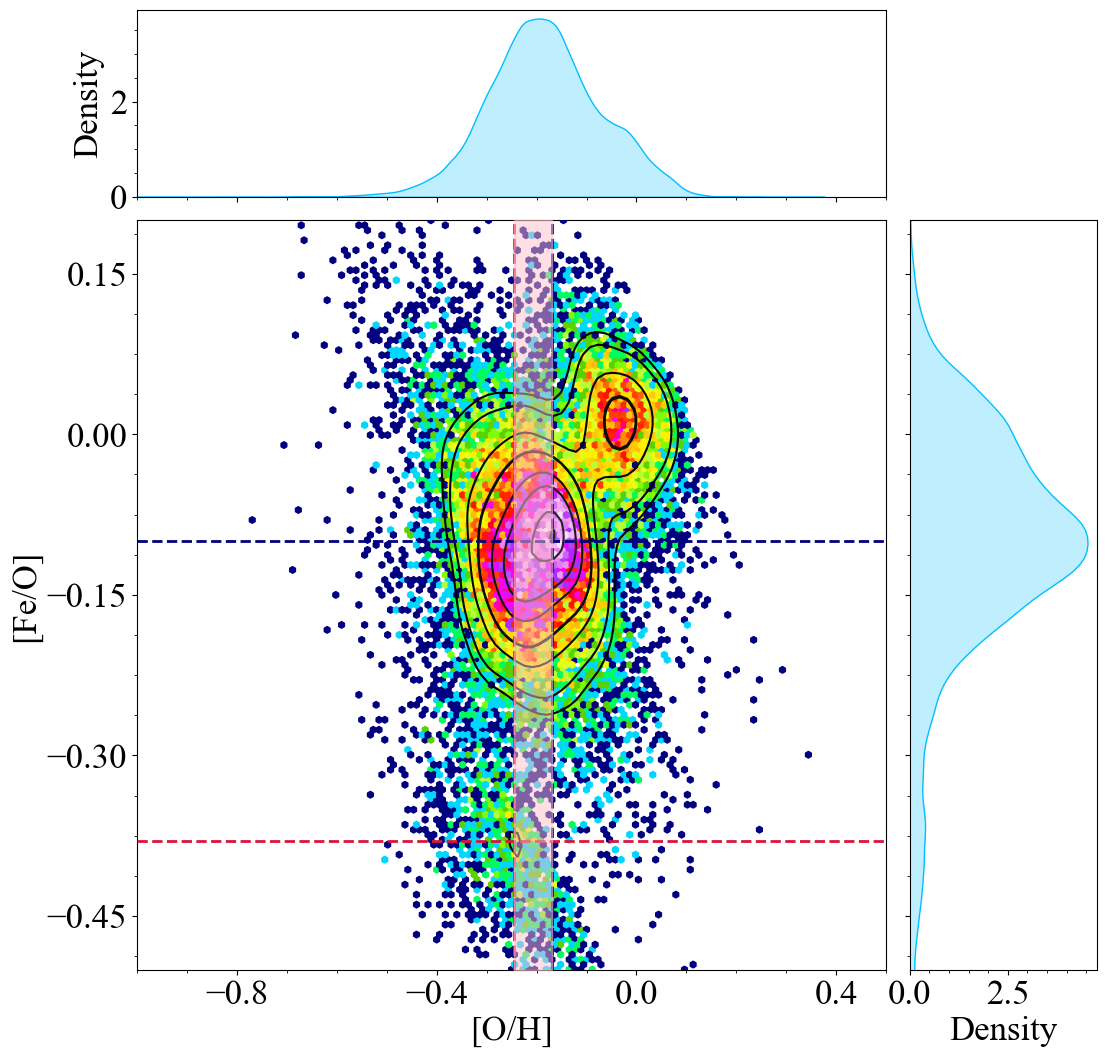}
  \caption{
  As in Fig. \ref{data_density}, but for APOGEE DR17 stars with  birth radii  between 7 and 9 kpc. The birth radii were derived in \citet{ratcliffe2023}, following the methodology given in
  Section \ref{sec_BR}.  In the two upper panels we also show with purple lines the predictions of  the fiducial chemical evolution model  of \citet{chen2023} computed at 8 kpc.  }
\label{data_density_BR}
\end{figure*}

\begin{figure*}
\centering
\includegraphics[scale=0.2]{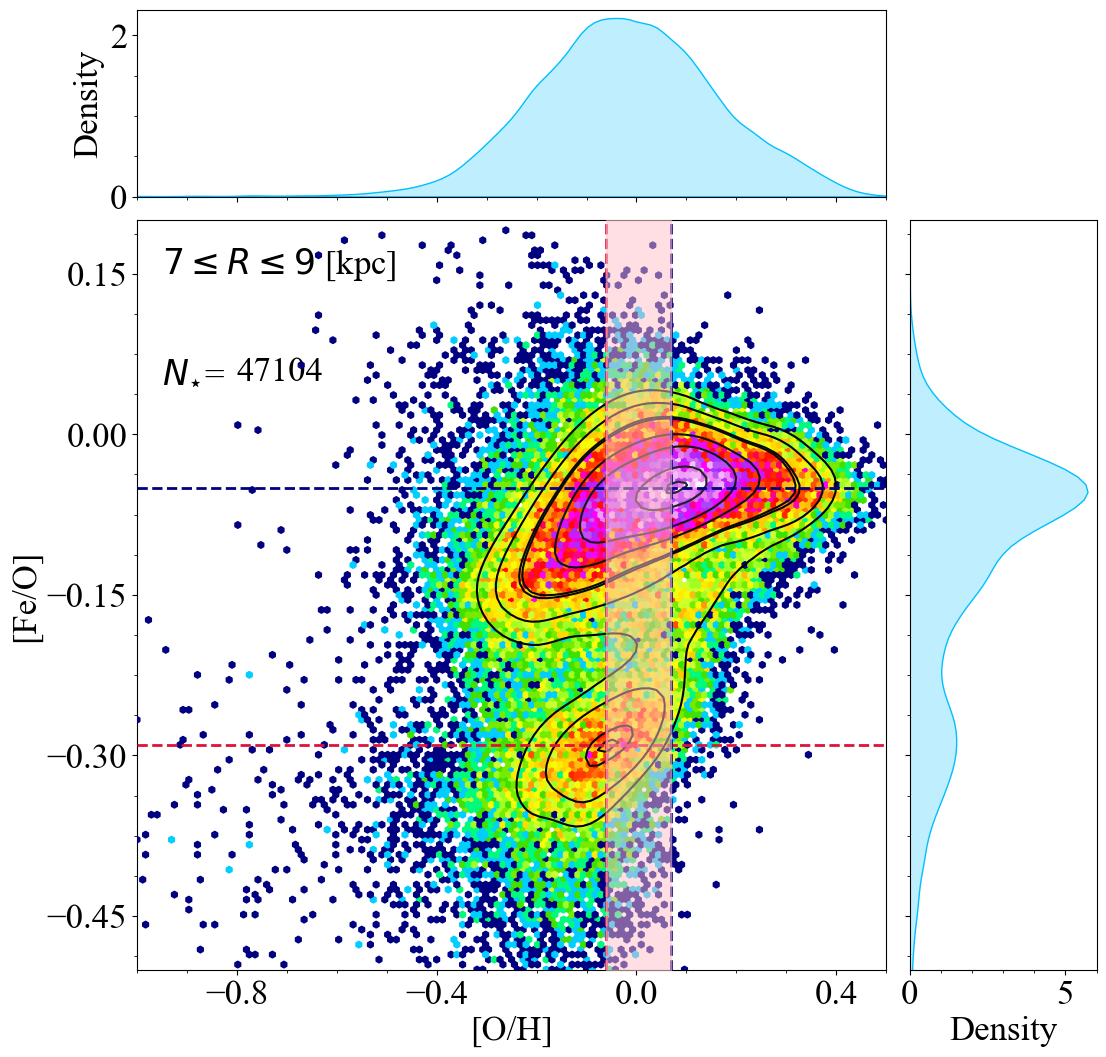}
\includegraphics[scale=0.2]{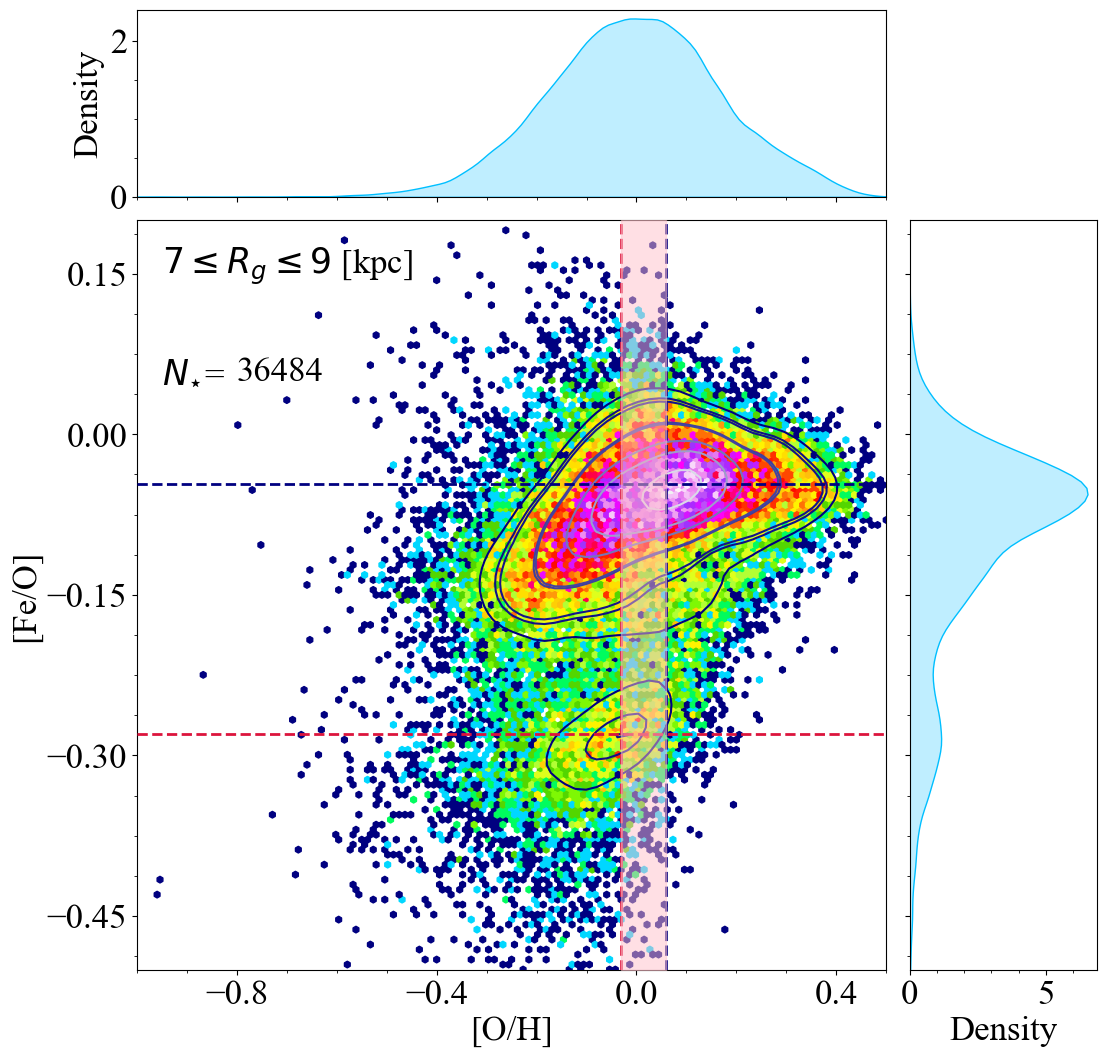}
\includegraphics[scale=0.2]{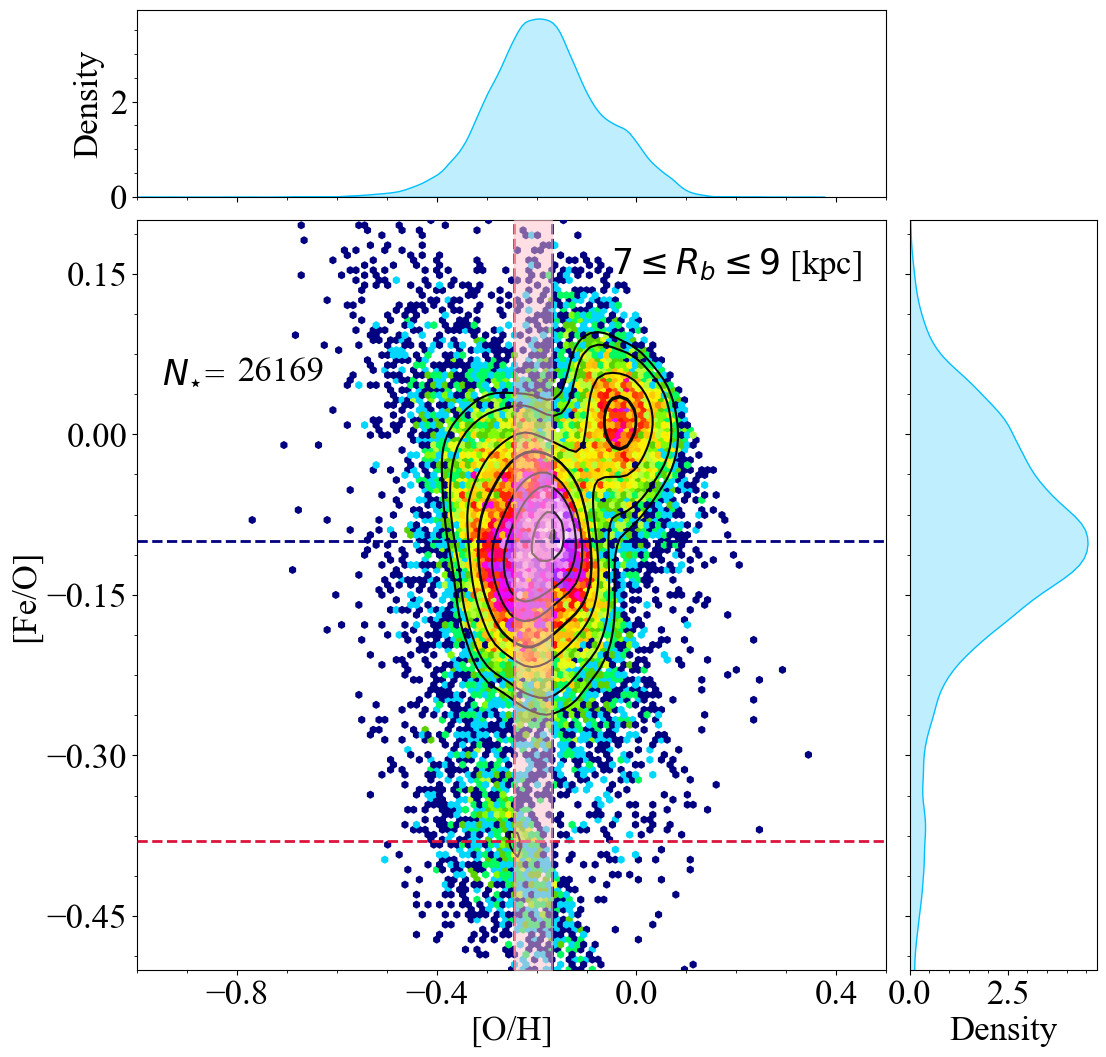}
  \caption{
   [Fe/O] versus [O/H] ratios for APOGEE DR17 are shown for the three different samples presented in this work: selected using Galactocentric distances (left panel), guiding radii (middle panel), and birth radii (right panel). Each panel indicates the number of stars in the sample. The colour maps and lines are consistent with those in Fig. \ref{data_density}. }
\label{comp_fig}
\end{figure*}

We define $\Delta$[Fe/H]$_{\rm peak}$ as the difference between the [Fe/H] values corresponding to the highest stellar density in the low-$\alpha$ and high-$\alpha$ sequences, the left panels of Fig. \ref{data_density} yield:
\begin{equation}
\Delta {\rm [Fe/H]}_{\rm peak}= {\rm [Fe/H]}_{{\rm peak} , {\rm low}} -{\rm [Fe/H]}_{{\rm peak} , {\rm high}} \simeq 0.42 \, \rm dex.
\label{eq:peak}
\end{equation}
If the reason for $\alpha$-bimodality is the star formation hiatus, the abundance ratios [Fe/$\alpha$] versus [$\alpha$/H] (right panels in Fig. \ref{data_density}) should exhibit a sharp increase of [Fe/$\alpha$] at nearly constant [$\alpha$/H] during the transition between the two disc phases, as discussed in \citealt{gratton1996}. Therefore, it is expected that $\Delta$[Fe/H]$_{\rm peak}$ >> $\Delta$[$\alpha$/H]$_{\rm peak}$.

Moreover, different $\alpha$ elements should display slightly different trends based on the nucleosynthesis progenitors. Elements exclusively synthesised by CC-SNe, such as oxygen, should precisely trace the SFR of the Galaxy, and a hiatus would lead to a sudden halt in the production of this chemical element.
In the right panels of Fig. \ref{data_density}, we highlight the region spanned by $\Delta$[$\alpha$/H]$_{\rm peak}$, revealing that $\Delta$[Mg/H]$_{\rm peak}=$ 0.18 dex, $\Delta$[Si/H]$_{\rm peak}=$ 0.22 dex, and $\Delta$[O/H]$_{\rm peak}=$ 0.13 dex. The smallest variation is observed for oxygen, consistent with the discussion above, while silicon exhibits the largest $\Delta$[$\alpha$/H]$_{\rm peak}$, given its substantial synthesis in Type Ia SNe (about 1/3, as indicated by \citealt{johnson_jenni_2020}).

\subsection{The APOGEE abundance distribution with birth radii}
\label{sec_BR}
 In Figs \ref{data_density} and \ref{guiding_radii}, we have  shown that the high- and low-$\alpha$ sequences made of stars, currently located in the solar neighborhood, are separated in the [Fe/$\alpha$]-[$\alpha$/H] plane, with a large difference in [Fe/$\alpha$] for only a small change in [$\alpha$/H]. However, the solar neighborhood is made of stars that were formed throughout the disc and have migrated there over time, as shown in simulations (e.g., \citealt{minchev2013}, \citealt{agertz2021},  \citealt{carrillo2023}). Chemical abundance trends have been shown to drastically change when using the stars guiding or current radii instead of their birth location   \citep{ratcliffe2023,ratcliffe2024}. Thus, to better understand the [Fe/a]-[$\alpha$/H] plane, we additionally show it using disc stars that were born in the solar neighborhood.

  \citet{ratcliffe2023,ratcliffe2024} estimated the birth radii of MW disc stars based on stellar ages and metallicities assuming that the gradient of metallicity, at birth, remains linear. Consequently, for any given lookback time ($\tau$), the birth-radius $R_b$ can be expressed as:
\begin{equation} 
\label{eqn:rb}
R_b(\text{age}, \text{[Fe/H]}) = \frac{\text{[Fe/H]} (R_b, \tau)  - \text{[Fe/H]}(R=0, \tau) }{\nabla \text{[Fe/H]}},
\end{equation}
where [Fe/H]$(R_b, \tau)$ is the metallicity at the lookback time  $\tau$,
$\nabla$[Fe/H]  the metallicity gradient at that time, and    [Fe/H]($R=0$, $\tau$) is the metallicity at the Galactic center.
 The time evolution of the metallicity gradient in the MW disc is able to be recovered directly from the data, as illustrated in \citet{ratcliffeTNG2024} for stronger-barred MW/M31-type galaxies from TNG50 simulation (see also \citealt{lu2022}). This fact allows us to investigate the [Fe/$\alpha$]-[$\alpha$/H] relationship for stars that have formed at a specific location in the galaxy with minimal modeling, giving better insight into the evolution of these elements. 

In Fig. \ref{data_density_BR}, the left panels show the [$\alpha$/Fe] versus [Fe/H]  ratios (for $\alpha$ = Mg, Si, O), while the right panels illustrate [Fe/$\alpha$] versus [$\alpha$/H], for APOGEE DR17 red giant disc stars with birth radii enclosed between 7 and 9 kpc. In the low-$\alpha$ sequence, two 
well-defined over-density clumps are visible, and \citet{ratcliffe2023} suggested that they could correlate with the triggered star formation due to the pericentric passages of Sagittarius galaxy \citep{lara2020}.

The primary focus of our study lies in the clear absence of stars between the high-$\alpha$ and the initial clump  within the low-$\alpha$ sequence (low metallicity). As detailed in Fig. \ref{data_density_BR}, our analysis concentrates into exploring various abundance ratios to discern potential signatures of a star formation hiatus. In agreement with Figs. \ref{data_density} and \ref{guiding_radii}, the plots depicting [Fe/$\alpha$] versus [$\alpha$/H] ratios (right panels in Fig. \ref{data_density_BR}) reveal a pronounced rise in [Fe/$\alpha$] at nearly constant [$\alpha$/H], for the considered $\alpha$ elements. In particular, for oxygen this bump is more pronounced. As discussed in Section \ref{peaks_obs}, this is a clear signature of a hiatus in the star formation history,  which is more pronounced when accounting for where stars were born.

It is worth noting that the chemical enrichment in the solar vicinity,  as inferred from the birth radii analysis presented in Fig. 
\ref{data_density_BR}, seems slightly different from the trends observed in Figs. \ref{data_density} and \ref{sel_gui}.
However, in the present paper, we will not show chemical evolution models tailored to reproduce the abundance ratios as constrained by  birth radii, because such a study is reserved for future work.

In the upper panels of Fig. \ref{data_density_BR}, we included the chemical evolution track of the fiducial model for the chemical evolution at 8 kpc as predicted by \citet{chen2023}. This model proposes that the APOGEE data in the solar vicinity can be explained by the various contributions to chemical enrichment from different radii.  The chemical evolution track computed at 8 kpc appears to trace the abundance data adequately. Moreover, in their Figure 9, \citet{chen2023} showed  a clear bimodal distribution in the predicted [$\alpha$/Fe] stellar distribution. However, it is worth mentioning that the distribution for stars with higher [$\alpha$/Fe] abundance ratios shows an offset when compared to APOGEE DR17 data. In addition, their age-metallicity relation for individual birth radii is inconsistent with the \citet{ratcliffe2023, ratcliffe2024} in that it is mostly flat for all radii at lookback time < 8 Gyr.

New multi-zone chemical evolution models need to be developed to accurately reproduce the abundance ratio distributions constrained by the birth radii of \citet{ratcliffe2023, ratcliffe2024}. In particular, the temporal evolution of the abundance gradients should be considered as a fundamental constraint.

 Finally, in Fig. \ref{comp_fig}, we compare the [Fe/O] versus [O/H] abundance ratios for the three different selections of APOGEE DR17 stars analyzed in this study: Galactocentric distances, guiding radii, and birth radii. We observe that the narrowest $\Delta$[O/H]${\rm peak}$ occurs for the birth radii case. As expected, the guiding radius selection accounts only for a portion of stellar migration \citep{minchev2018}, resulting in an intermediate $\Delta$[O/H]${\rm peak}$ value compared to the other two cases analysed.

\section{Chemical evolution models }
\label{models_sec}
In this Section, we present the chemical evolution models adopted in this work.  In particular, we developed a model for the MW and a model for dwarf galaxies, which could have been accreted by the MW in different phases of its evolution. 
This particular choice is  also   motivated by the fact that  previous chemical evolution models  
where only  gas infall with primordial chemical composition  was taken into account \citep[i.e.][left panel of their Figure 16]{spitoni2021}, were not able to reproduce APOGEE DR16 distribution of [Mg/Fe] in the high-$\alpha$ sequence. 
In Sections \ref{2infall_sec} and \ref{dwarf_sec},  we report the main properties of  MW discs and  dwarf models, respectively. In Section   \ref{nucleo_sec}, the nucleosynthesis prescriptions will be discussed.

\subsection{The two-infall model for the MW disc components}
\label{2infall_sec}
 
In this paper, we  use a similar  two-infall chemical evolution model, as presented in  \citet{nissen2020}, 
to reproduce the abundance ratios and the age metallicity relation observed in stellar HARPS spectra. 
The functional form of the adopted gas infall rate is:
\begin{eqnarray}
\mathcal{I}_i(t)&=& X_{i} \, \Big( \overbrace{A \, e^{-t/ {\rm T}_{ \rm high}}}^{\text{\textcolor{red}{{1st infall, high-$\alpha$}}}}+ 
 \overbrace{\theta(t-{\rm T}_{{\rm max}})  \, B \, e^{-(t-{\rm T}_{{\rm max}})/{\rm T}_{ \rm low}}}^{\text{\textcolor{blue}{{2nd infall, low-$\alpha$}}}} \Big),
\label{infall}
 \end{eqnarray}
where ${\rm T}_{{\rm high}}$, ${\rm T}_{{\rm low}}$,     are the timescales of the two distinct gas infall episodes. 
The Heaviside step function is represented by $\theta$. 
  $X_i$ is the abundance by mass unit of the element $i$ in the
infalling gas. 
Finally, the coefficients $A$ and $B$ are obtained by imposing a fit to the observed current total surface mass density in the solar neighbourhood. Following \citet{spitoni2020}, we use the value of total surface mass density in the solar neighbourhood of 47.1 $\pm$ 3.4 M$_{\odot} \rm{ pc}^{-2}$ as provided by \citet{mckee2015}. We adopt the \citet{scalo1986} initial stellar mass function (IMF), constant in time and space.

 Following the model in \citet{nissen2020},  we assume that the high-$\alpha$ phase starts at [Fe/H]=-0.8 dex,  as a level of pre-enrichment (see Section \ref{dwarf_sec}).
The quantity ${\rm T}_{{\rm max}}$ is the time of the maximum infall rate in the second accretion episode, i.e. it indicates the delay between the two peaks of infall rates.  The SFR is expressed as the \citet{kenni1998} law,
\begin{equation}
\psi(t)\propto \nu_{{\rm high}, {\rm low}} \cdot \sigma_{g}(t)^{k},
\label{k1}
\end{equation}
 where $\sigma_g$ is the gas surface
 density and $k = 1.5$ is the exponent. 
The quantity $\nu_{{\rm high}, {\rm low}} $ is the star formation efficiency (SFE), which can be different in different Galactic evolutionary phases. 
In Table \ref{tab_A}, the values adopted for the parameters introduced above are reported. We stress that we assume the same values as \citet{nissen2020} for  the ${\rm T}_{{\rm high}}$, ${\rm T}_{{\rm low}}$, ${\rm T}_{{\rm max}}$,   ${\nu}_{{\rm high}}$ and ${\nu}_{{\rm low}}$ parameters.
To reproduce the distribution between high- and low-$\alpha$ stars (see Section \ref{count_stars}) we impose that the ratio of the total surface mass densities of the thick (high-$\alpha$) and thin (low-$\alpha$) discs is $\sigma_{\rm{low}}$/$\sigma_{\rm{high}}=2.5$, a smaller value compared to previous works of \citet{spitoni2020,spitoni2021}.
 However, it is still compatible with  \citet{fuhr2017} work where the author derived a local mass ratio as low as 1.73 after correction for the  difference in the scale height.

\subsection{The chemical evolution model for a massive dwarf galaxy}
\label{dwarf_sec}
\begin{figure}
\centering
 \includegraphics[scale=0.28]{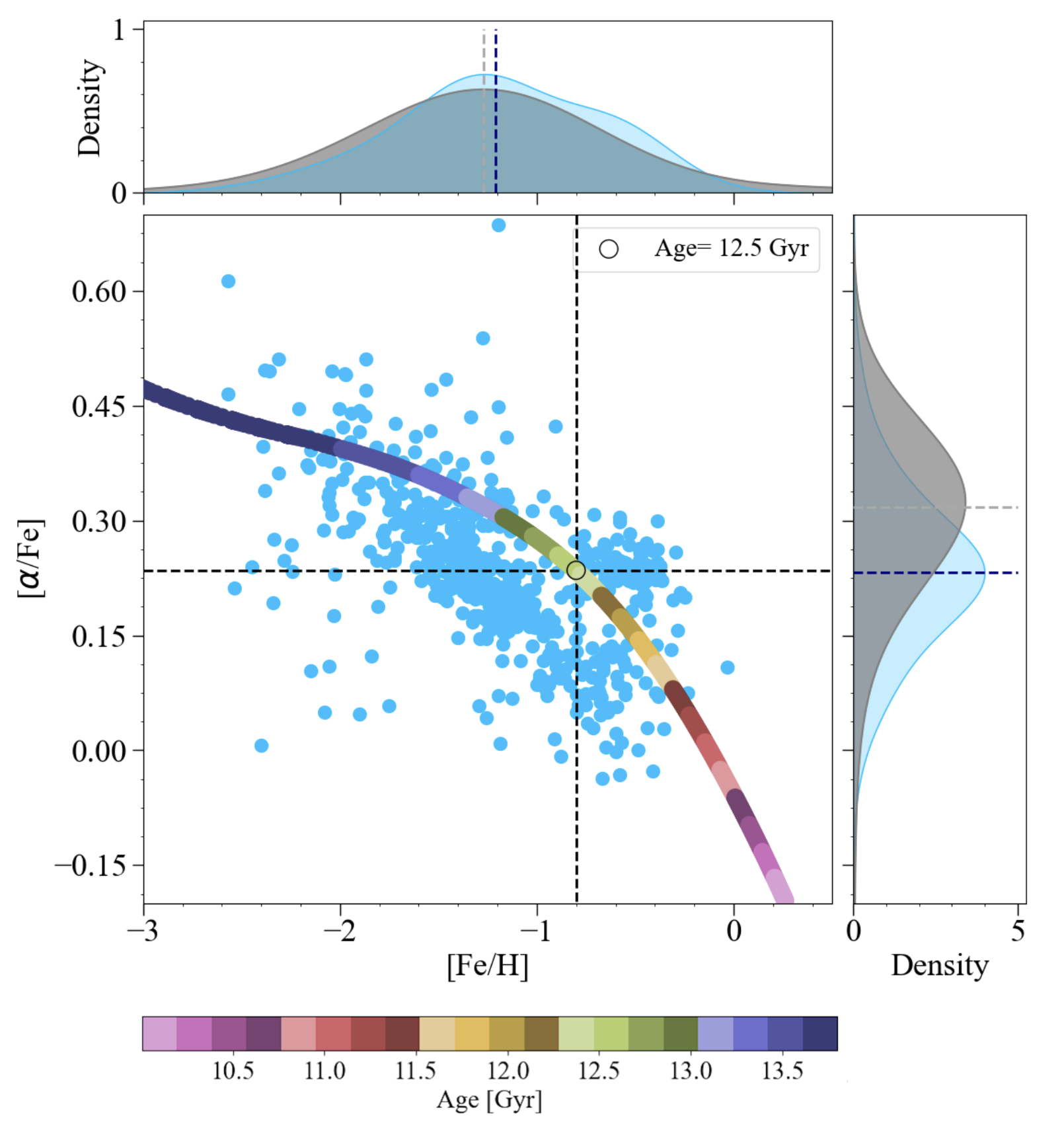}
  \caption{[$\alpha$/Fe] versus [Fe/H] predicted by the accreted dwarf satellite presented in Section \ref{dwarf_sec}. The coloured coding stands for the galactic ages at different evolutionary phases. With the empty circle, the chemical abundance ratio at an age of 12.5 Gyr is reported. With light-blue points, we labelled Enceladus stars as suggested by \citet{helmi2018}. On the sides of each panel, the observed (light-blue shaded area) and predicted (dark-grey shaded area) normalised KDE
of the abundance ratio distributions calculated with a Gaussian kernel are also reported. }
\label{dwarf_fig}
\end{figure}

 As mentioned in Section \ref{2infall_sec}, we assume that  the Galactic disc has been built up by already chemically enriched gas and that the high-$\alpha$ phase starts at [Fe/H]=-0.8 dex. The chemical enrichment in the [$\alpha$/Fe] versus [Fe/H] (where for $\alpha$ we mean Si+Mg)  required to reproduce the main properties of the high-$\alpha$ sequence APOGEE DR17 red giant stars  (i.e. the  [$\alpha$/Fe] distribution, see Section \ref{results_sec}) is reported  with an empty circle in Fig. \ref{dwarf_fig}.
This level of chemical enrichment can be reached by  a "typical"  dwarf galaxy that can be accreted after an evolutionary time of 1.3 Gyr (12.5 Gyr ago). 
We considered for such a galaxy a 1-infall model with an infall time-scale of 0.24 Gyr  and  the SFE fixed to 0.42 Gyr$^{-1}$ \citep{vincenzo2019}.
We assume the presence of a galactic wind  (common in dwarf galaxies) proportional to the SFR:
\begin{equation}
    W(t)=\omega \, \psi(t),
\end{equation}
where $\omega$ is the loading factor fixed to the value of 1.5.
The SFR is defined by the  \citet{kenni1998} law. 
 In Fig. \ref{dwarf_fig} we report  the chemical evolution  in the [$\alpha$/Fe] versus [Fe/H] (where for $\alpha$ we mean Si+Mg)
abundance ratio space predicted by the dwarf model compared with \citet{helmi2018} data for the Enceladus system. 
Our objective in this paper is not to precisely reproduce the data of Enceladus, but rather to closely approximate the Metallicity Distribution Function (MDF).
In Fig. \ref{dwarf_fig} we highlight the chemical composition of the single stellar population (SSP) formed after 1.3 Gyr of evolution, corresponding to a Galactic age of 12.5 Gyr.
At this time our dwarf galaxy exhibits a surface mass density of 1.50 M$_{\odot} \, \rm{ pc}^{-2}$, a lower value compared to the 2 M$_{\odot} \, \rm{ pc}^{-2}$ value, as  predicted by   \citet{cescutti2020}.

 In conclusion, in this study, we assume that the chemical composition $X_i$ of the element $i$-th in Eq. (\ref{infall})  of the infalling gas is enriched with the chemical composition of a "typical" massive dwarf galaxy that can be accreted after an evolutionary time of 1.3 Gyr (12.5 Gyr ago). 
It is important to stress that we are not implying that the disc has been formed out of the accreted gas and stars by a single dwarf galaxy, but we try to justify the required level of pre-enrichment for the infalling gas.
This accreted gas, likely originated from filaments of intergalactic medium, should be contaminated by various dwarf galaxies. Therefore, the chemical pre-enrichment level of the infalling gas assumed in this study should be viewed as an upper limit, as some degree of dilution is expected to have indeed occurred. It is worth noting that in the VINTERGATAN chemo-dynamical simulation of a MW like Galaxy in the cosmological context \citep{agertz2021,renaud2021},   the thick disc - kinematically hot disc populated by high-[$\alpha$/Fe] stars - is formed both in situ and in accreted satellite galaxies (see Figure 8 in \citealt{renaud2021}). 

 Throughout the remainder of this article, we assume that the infalling gas during the thick and thin disc phases has the chemical composition predicted for the accreted dwarf galaxy after 1.3 Gyr of evolution.

\begin{table*}
\begin{center}
\tiny
\caption{Summary of the main parameters of the two-infall model  presented in this study (see Section \ref{2infall_sec}): the accretion timescales (T$_{\rm{high}}$, T$_{\rm{low}}$),  time-delay  T$_{{\rm max}}$, and the star formation efficiencies ($\nu_{\rm{low}}$,$\nu_{\rm{high}}$, 
the present-day total surface mass density ratios $\sigma_{\rm{low}}$/ $\sigma_{\rm{high}}$). Finally, the infall chemical composition   is reported indicating the  abundances by mass for O, Mg, Si and Fe in the last four columns. 
We also specify the values corresponding to those adopted in the revised version  of the \citet{spitoni2019} two-infall chemical evolution model presented in \citet{nissen2020}.}
\label{tab_A}
\begin{tabular}{|c|ccccc|c|cccc|}
\hline
  \hline
 & & & & & && &&&\\
Model MW  &    $T_{\rm{high}}$& $T_{\rm{low}}$ &
$T_{\rm max}$&
  $\nu_{\rm{high}}$ &  $\nu_{\rm{low}}$ &$\sigma_{\rm{low}}$/$\sigma_{\rm{high}}$& \multicolumn{4}{|c|}{$X_i$} 
\\
 &[Gyr]& [Gyr]&[Gyr] & [Gyr$^{-1}$]&[Gyr$^{-1}$] & &    \multicolumn{4}{|c|}{Enriched (Section \ref{dwarf_sec})} \\
 \hline
 & & & & & && &&&\\
  & & & & & && O&Mg&Si&Fe\\
  & 0.377& 3.203 & 3.519& 2.000& 0.600&2.500 & 1.573 $\cdot 10^{-1}$ &1.759 $ \cdot 10^{-2}$&  1.708 $\cdot 10^{-2}$& 1.859 $ \cdot  10^{-2}$ \\
 
  \hline 
& \multicolumn{5}{|c|}{\citet{nissen2020}} & & && &\\
 \hline
\end{tabular}
\end{center}
\end{table*}

\subsection{Nucleosynthesis prescriptions}
\label{nucleo_sec}
As for the models of \citet{spitoni2019} and \citet{nissen2020}, we employed the nucleosynthesis prescriptions introduced by \citet{francois2004} for Fe, Mg, Si, and O. Specifically, the Mg yields for massive stars from \citet{WW1995} were artificially increased to match the solar Mg abundance. Yields for stars in the mass range of 11–20 M$_{\odot}$ were augmented by a factor of seven, while those for stars with a mass larger than 20 M$_{\odot}$ were, on average, twice as large. No adjustments were necessary for the yields of Fe computed for solar chemical composition. For Si, only the yields of extremely massive stars (M > 40 M$_{\odot}$) were amplified by a factor of two.

In the case of O, adopting the original \citet{WW1995} yields from massive stars as functions of metallicity yielded the best agreement with the [O/Fe] vs. [Fe/H] relation and the solar O abundance, as determined by \citet{asplund2005}.
To maintain the observed [Mg/Fe] vs. [Fe/H] pattern, the yields of \citet{iwamoto1999} for Mg were increased by a factor of five. This set of yields has been widely utilised in the literature, acknowledging the inherent uncertainty in stellar yields as a component of chemical evolution models \citep[e.g.][]{francois2004,romano2010,cote2017,prantzos2018}.

The aforementioned set of yields, extensively employed in the literature \citep{cescutti2007,cescutti2022,spitoni2017,spitoni2D2018,spitoni_2D_2023, mott2013, vincenzo2019,palla2022}, has proven capable of replicating the key characteristics of the solar neighbourhood.

\section{Results}
\label{results_sec}
In this  Section, we present model results for the disc components using the chemical evolution model presented in Section \ref{2infall_sec}. Particularly, in  Section \ref{comp}, the comparison with APOGEE data will be presented. 
In Section \ref{model_stop}, we will discuss if  signatures of the hiatus in the SFR in the predicted abundance ratios are consistent with APOGEE observations.
 In Section \ref{count_stars}, we show  the distributions of predicted  stars in different regions of the  [Mg/Fe] versus [Fe/H] plane. 
Finally, in Section \ref{other_obs} we prove that our model is capable of reproducing other Galactic disc observables.

\subsection{Comparison with APOGEE DR17 data}
\label{comp}
\begin{figure*}
\centering
 \includegraphics[scale=0.24]{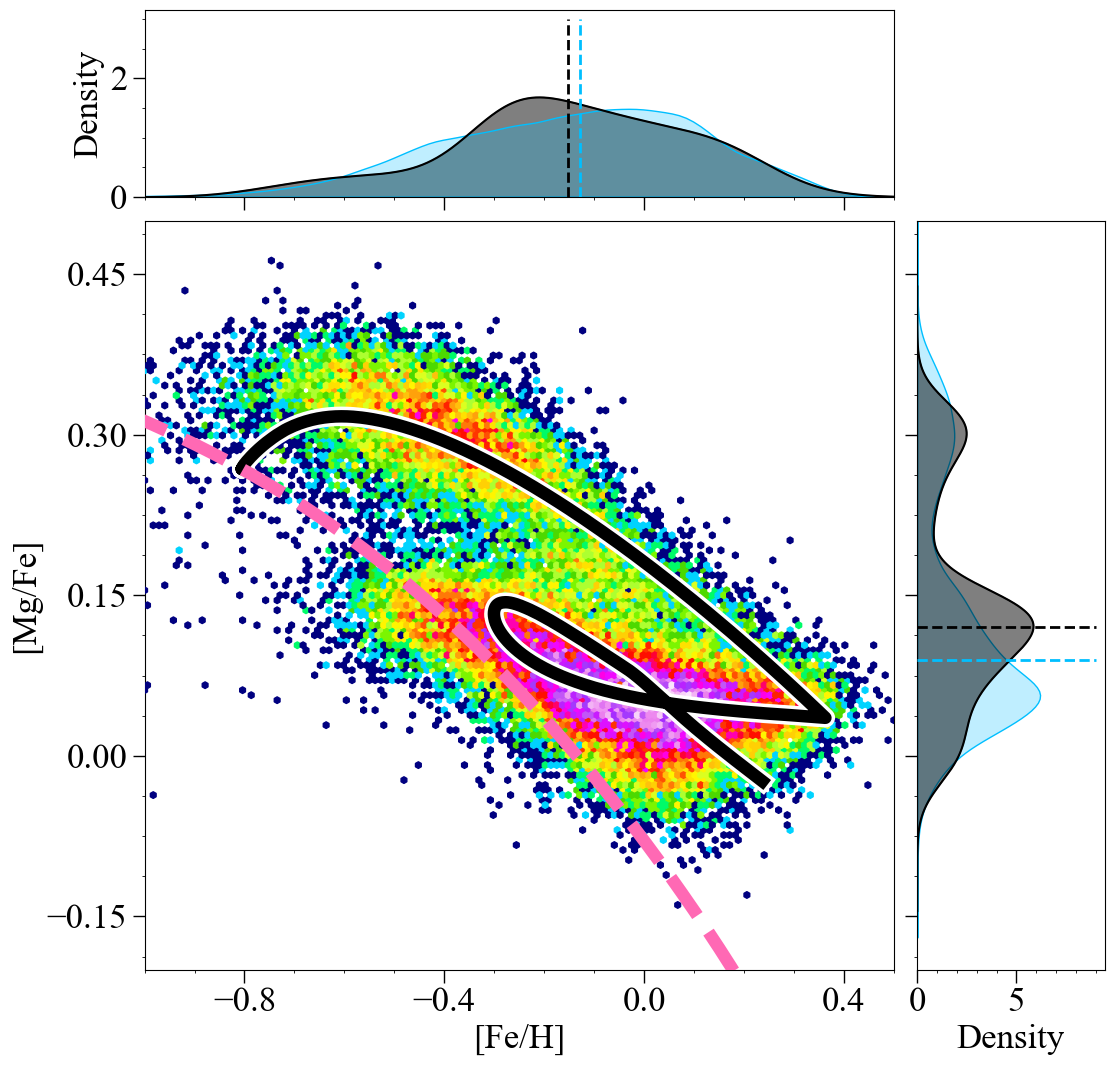}
 \includegraphics[scale=0.24]{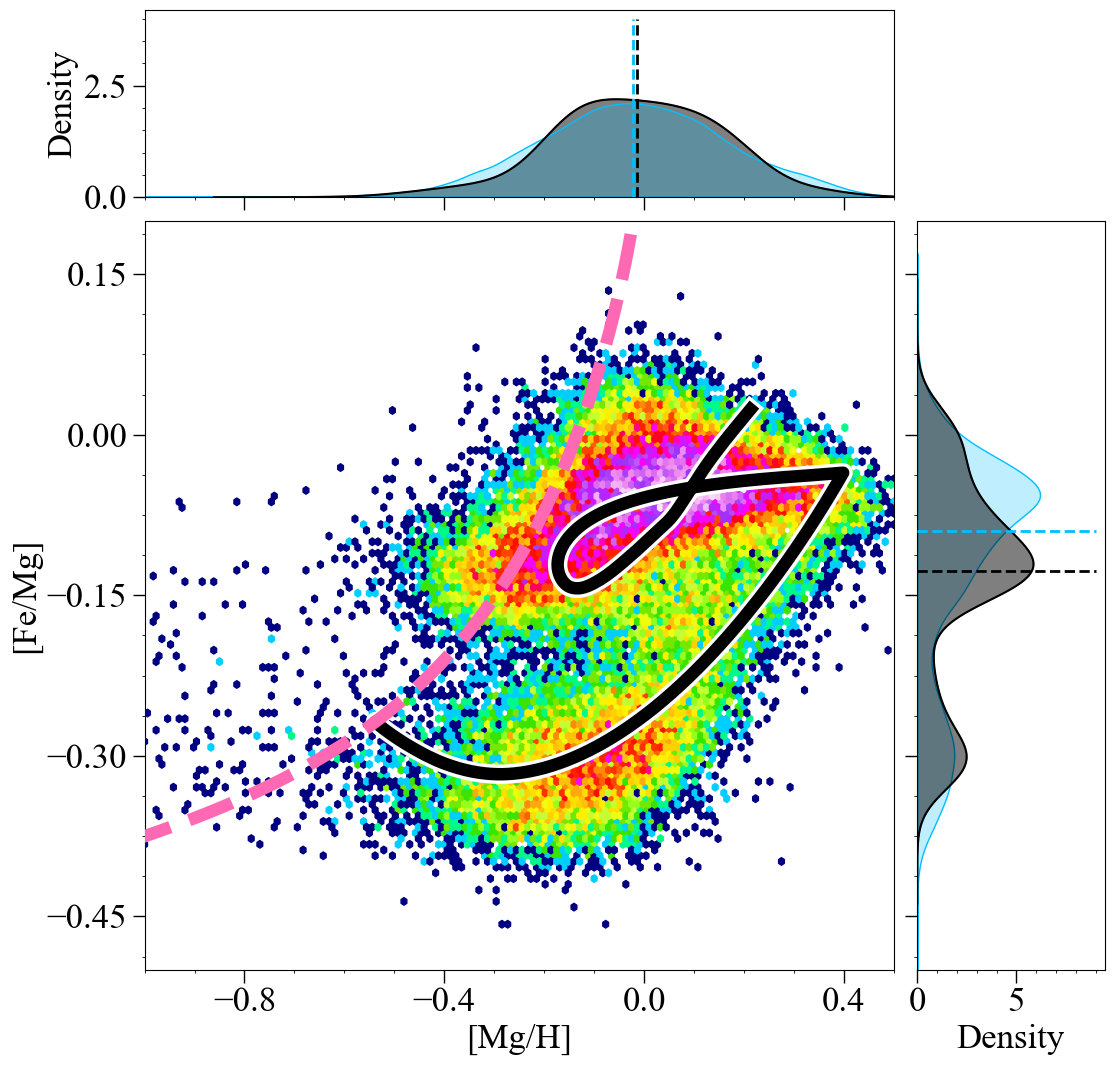}
  \includegraphics[scale=0.24]{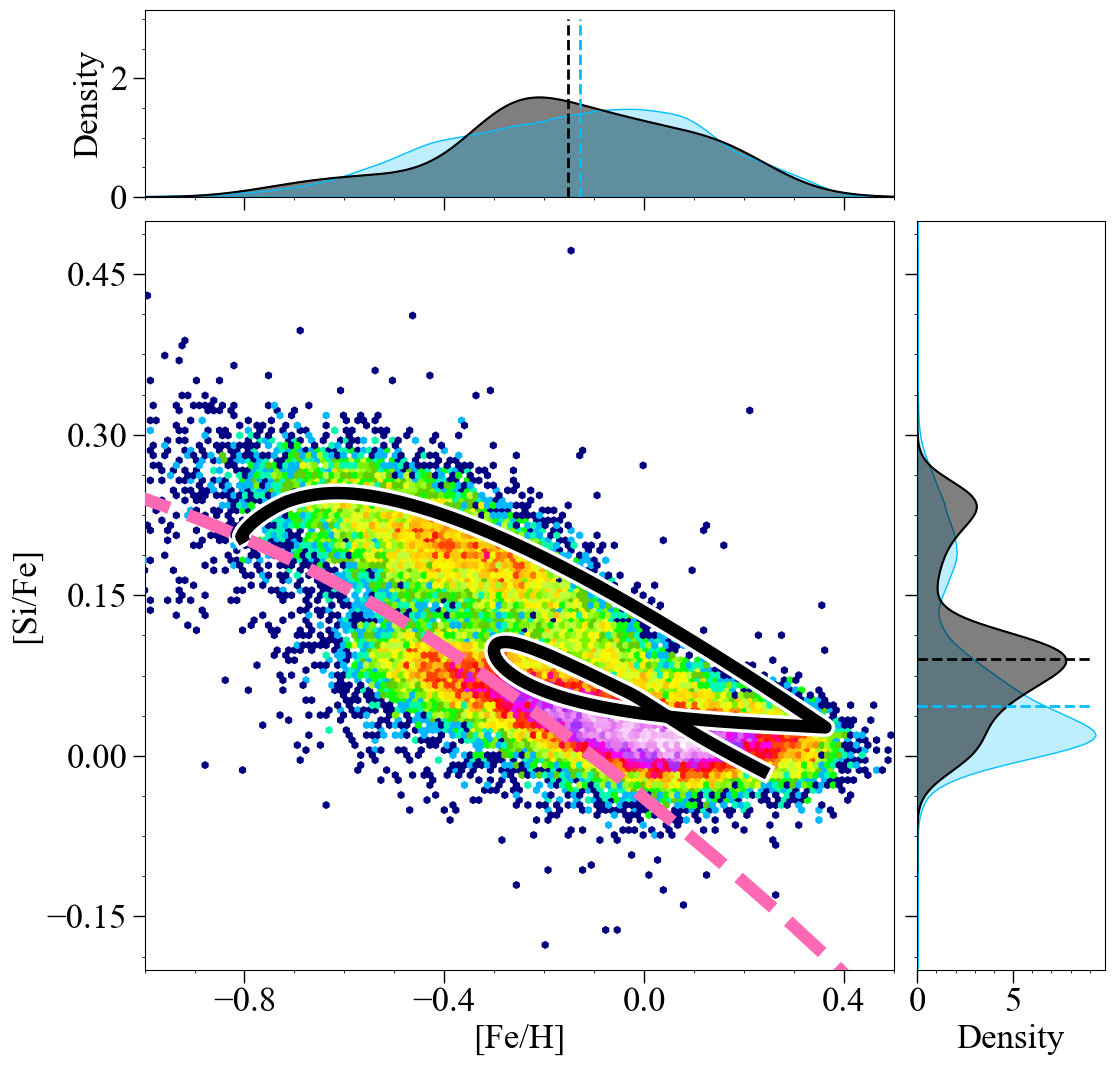}
   \includegraphics[scale=0.24]{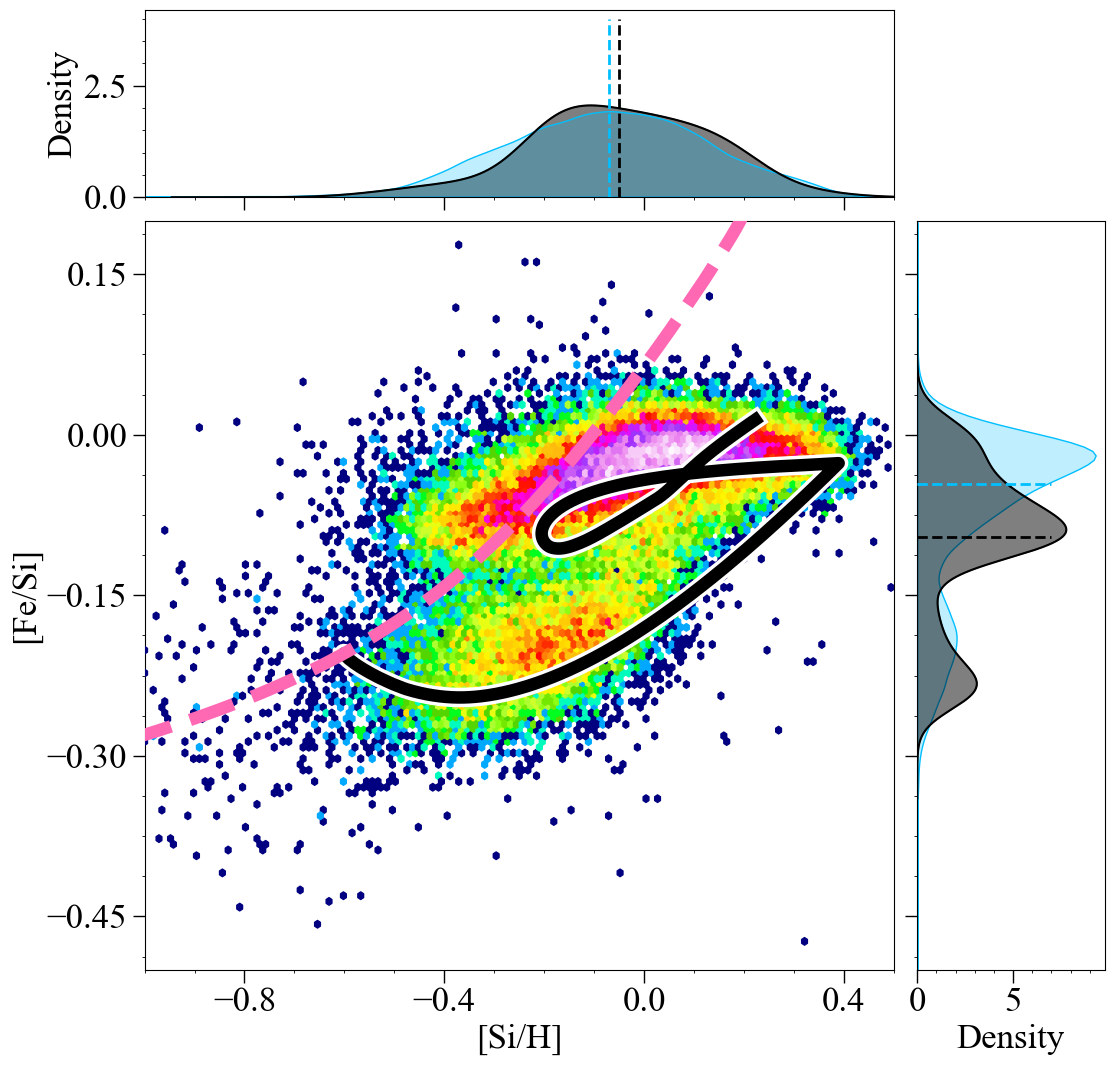}
 \includegraphics[scale=0.24]{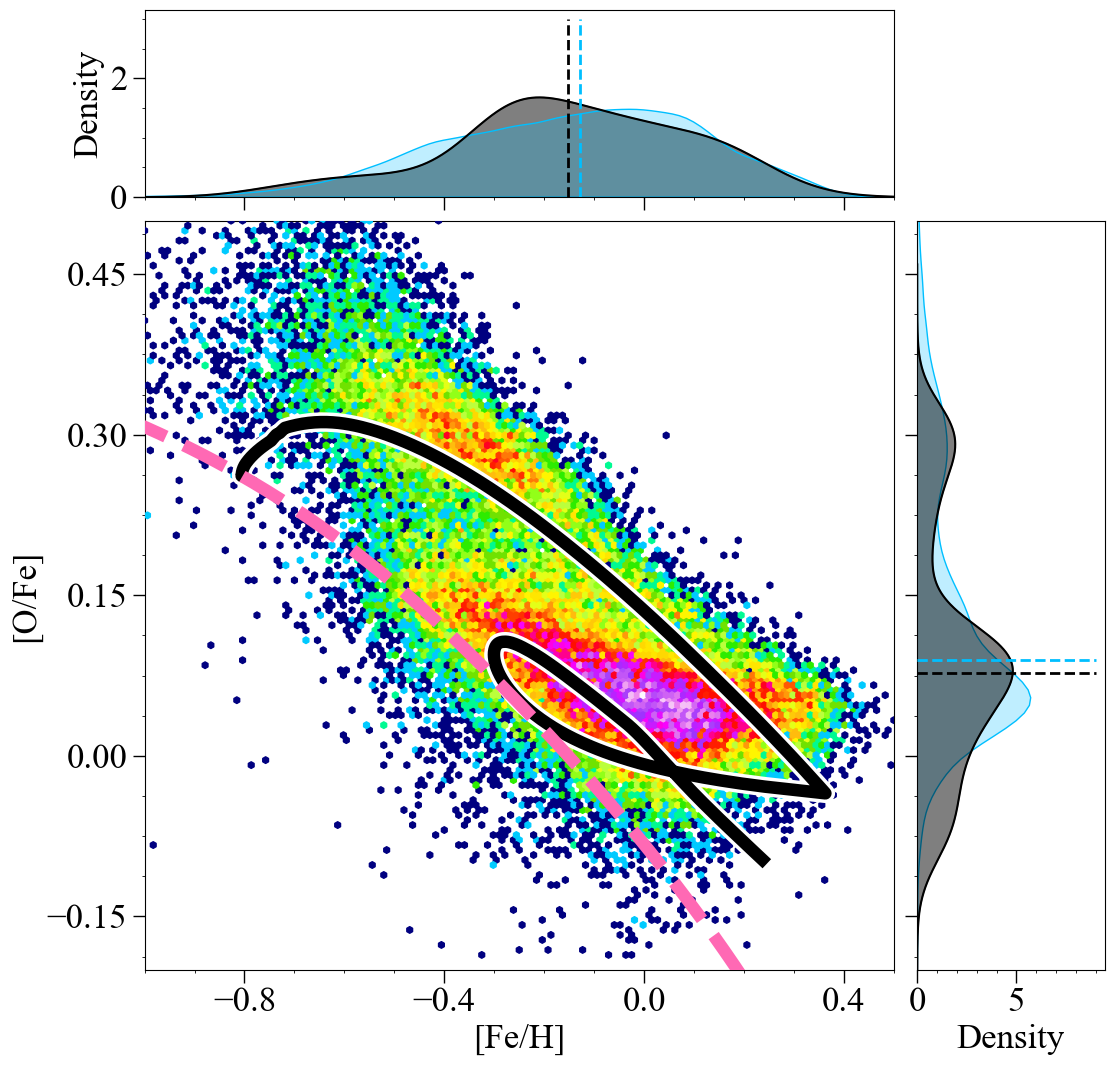}
 \includegraphics[scale=0.24]{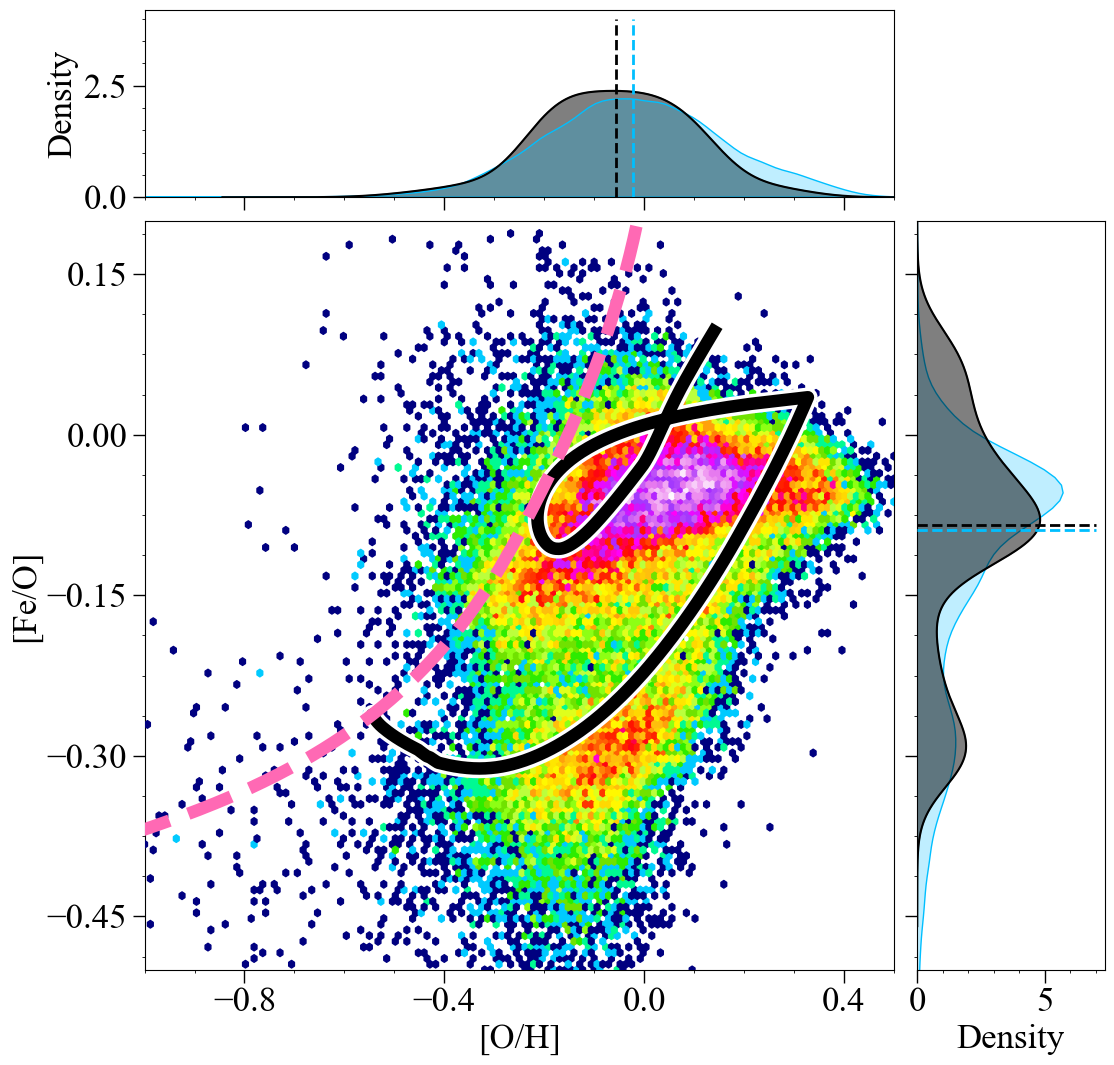}
  \caption{Comparison between predicted (black solid lines, see Section \ref{2infall_sec} for the two-infall model description) and observed APOGEE DR17 (density maps in logarithmic scale) stellar distribution of stars in the [$\alpha$/Fe] versus [Fe/H] (left panels) and [Fe/$\alpha$] versus [$\alpha$/H] (right panels) planes for $\alpha$ = Mg (upper panels), $\alpha$ = Si (middle panels) and  $\alpha$ = O (lower panels). Model predictions of the massive dwarf galaxy (see Section \ref{dwarf_sec} for model details) are reported with the   pink   dashed lines.  On the sides of each panel the observed (light-blue shaded area) and predicted (dark-grey shaded area) normalised KDE of the abundance
ratio distributions calculated with a Gaussian kernel are also reported. Finally with vertical lines are reported the median values of the respective distributions.  
  }
\label{fig_Mg_Si_O_model}
\end{figure*}

In Fig. \ref{fig_Mg_Si_O_model}, we compare model predictions for [$\alpha$/Fe] versus [Fe/H] (left panels) and [Fe/$\alpha$] versus [$\alpha$/H] (right panels) where $\alpha$ = Mg, Si, O with the APOGEE DR17 stars of Fig. \ref{data_density} adopting the solar  abundances of \citet{grevesse2007} to be consistent with the data.
The distribution of the stars as a function of the considered abundance ratios is also compared on the sides of each panel.
It is worth noting that the inclusion of enriched gas inflow, as detailed in Section \ref{dwarf_sec}, enables the model to accurately reproduce the distribution in [$\alpha$/Fe] during the high-$\alpha$ phase. This improvement highlights the limitations  that were present in previous chemical evolution models, (e.g. \citealt{spitoni2021} and \citealt{spitoni2019}), where the APOGEE DR16 distribution of [Mg/Fe] was not reproduced as well as it is in the present paper.

\begin{figure*}
\centering
\includegraphics[scale=0.24]
  {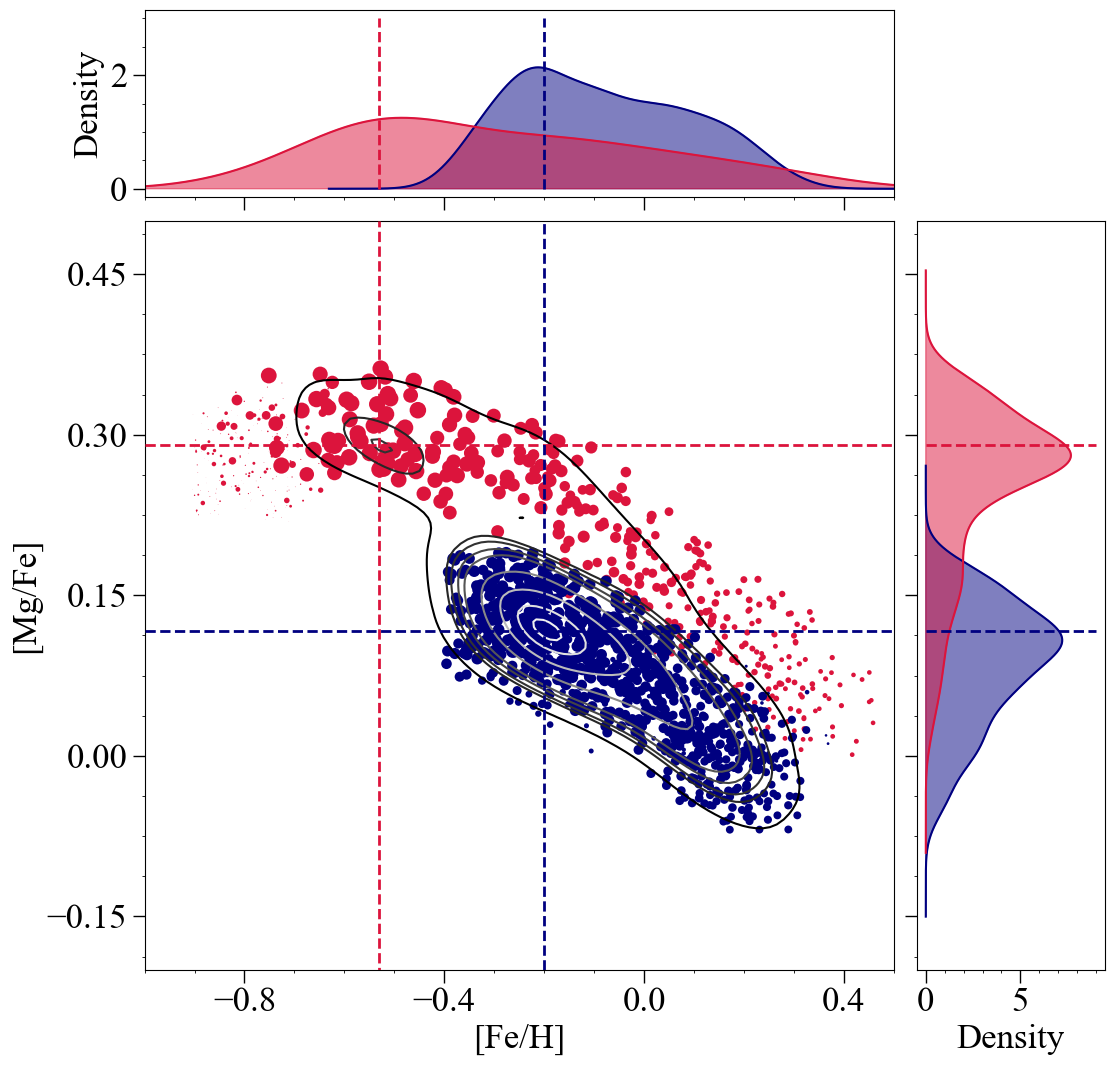}
  \includegraphics[scale=0.24]{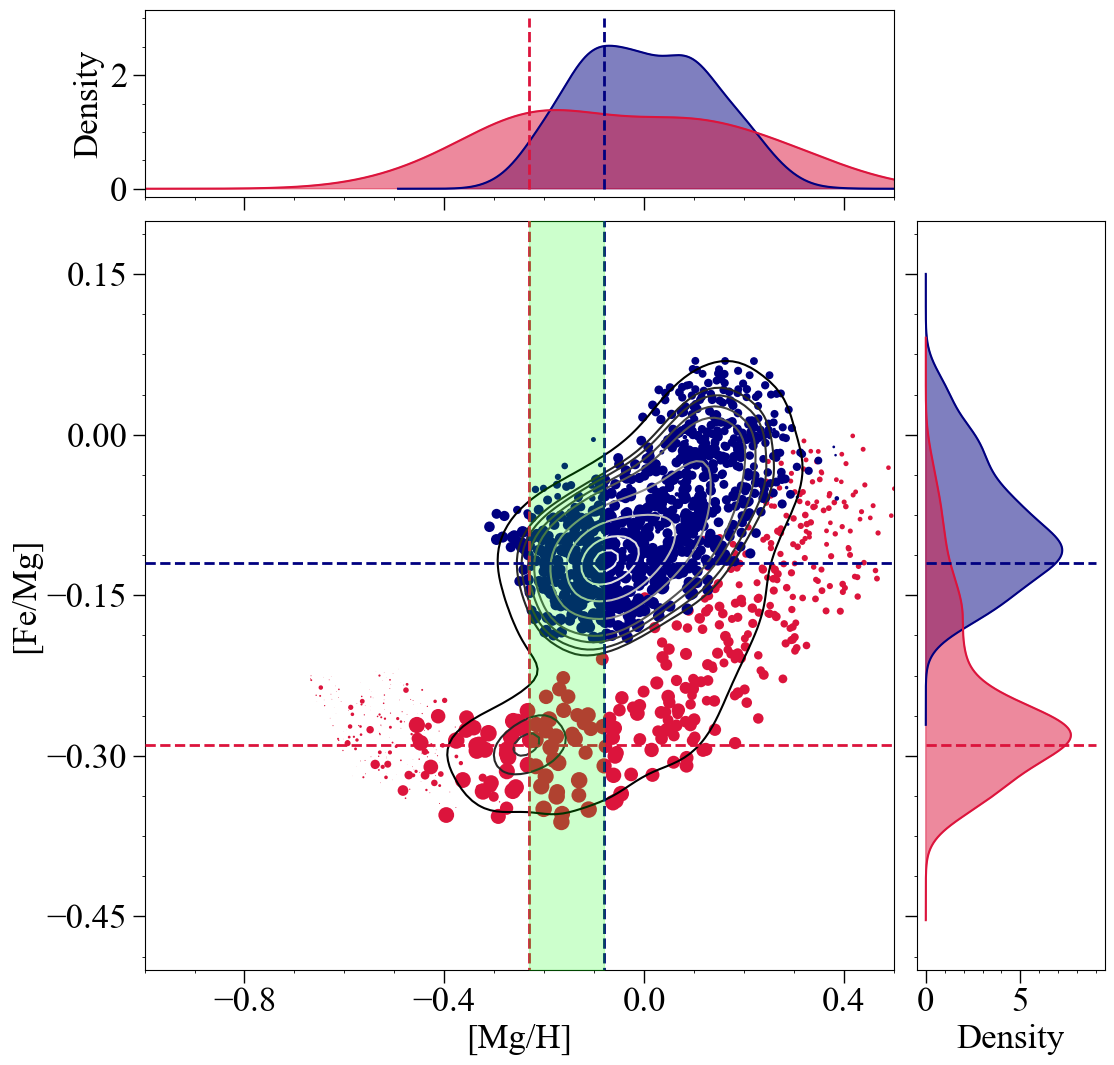}
  \includegraphics[scale=0.24]
   {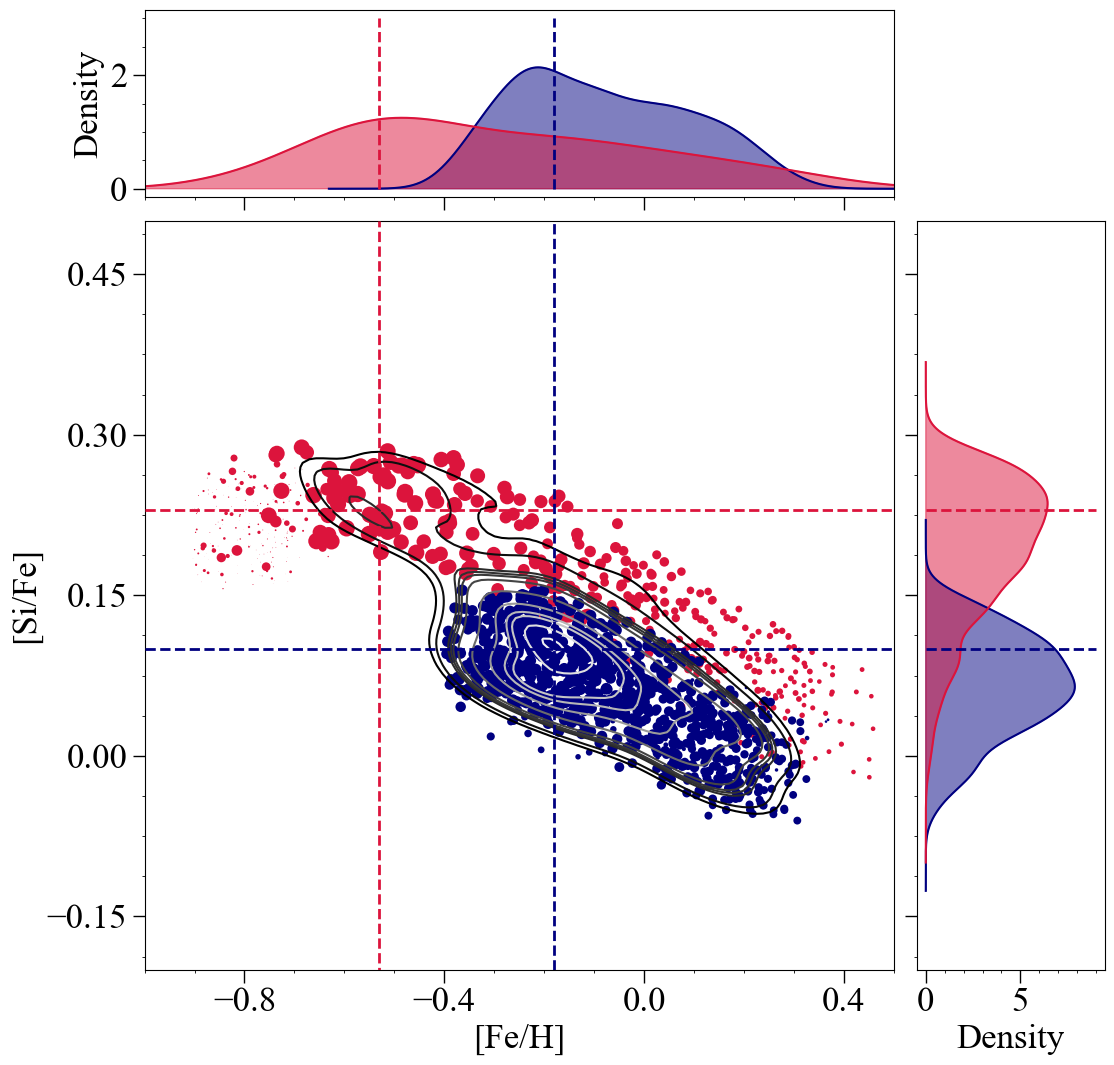}
 \includegraphics[scale=0.24]{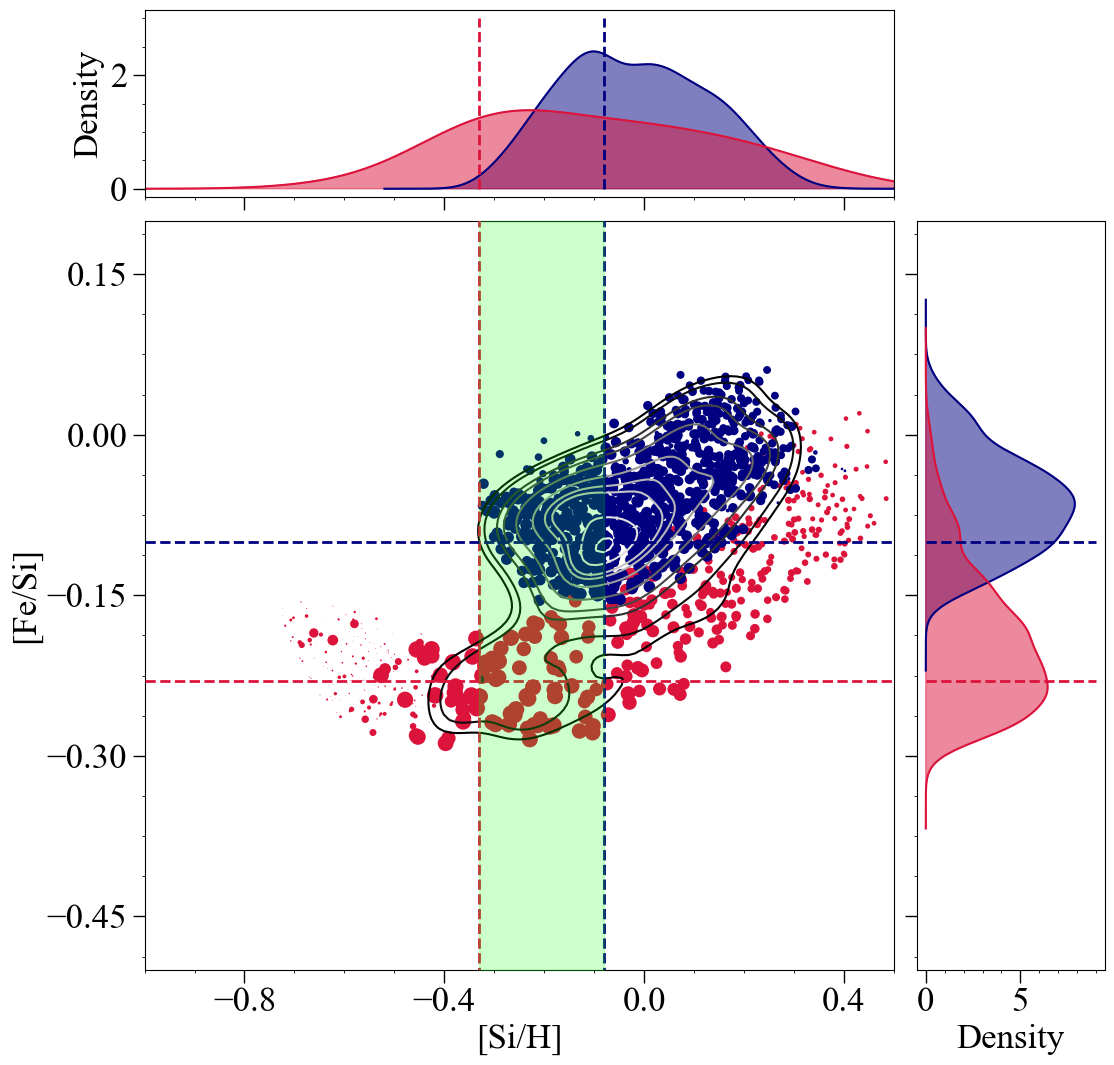}
  \includegraphics[scale=0.24]{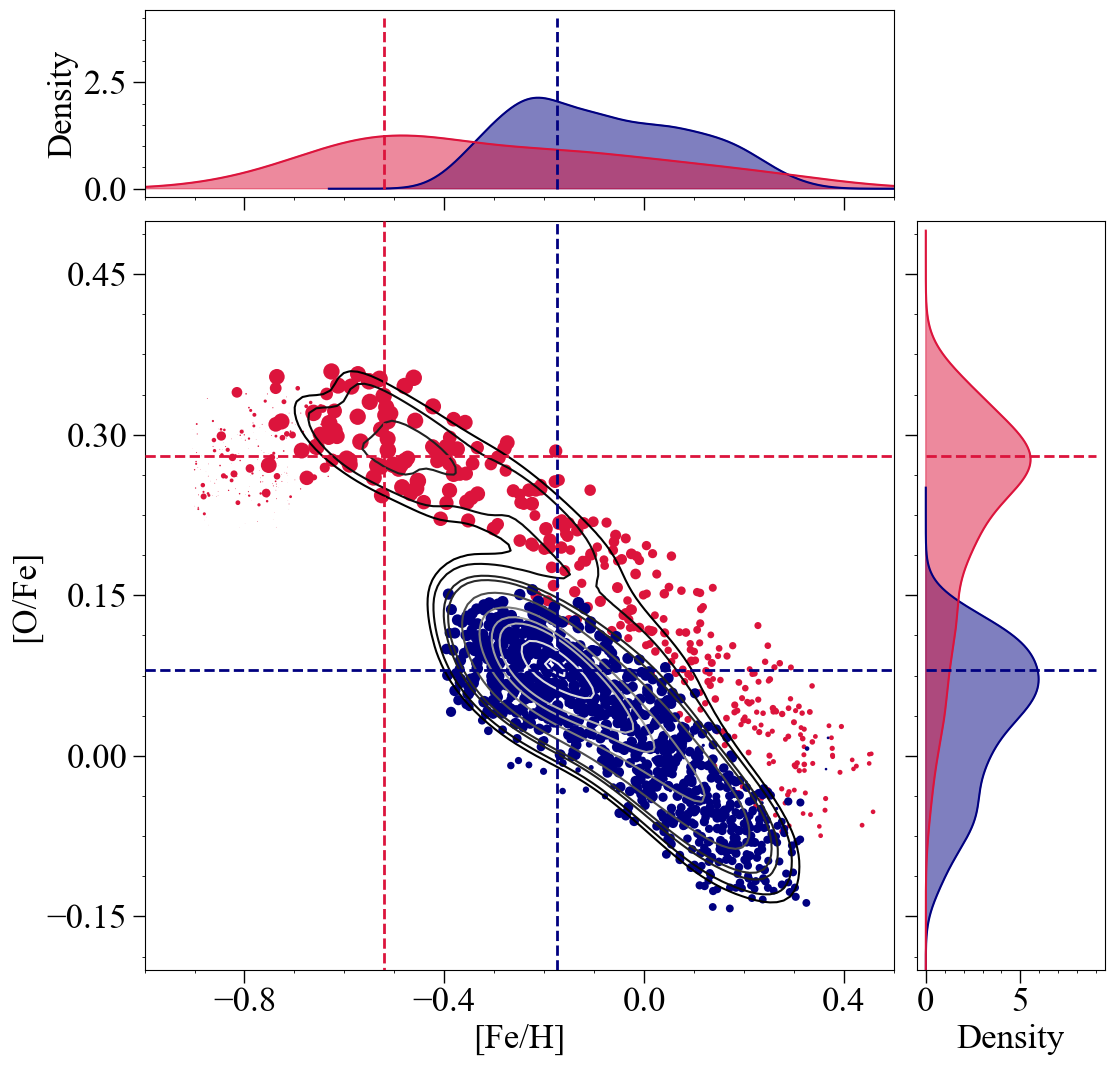}
 \includegraphics[scale=0.24]
 {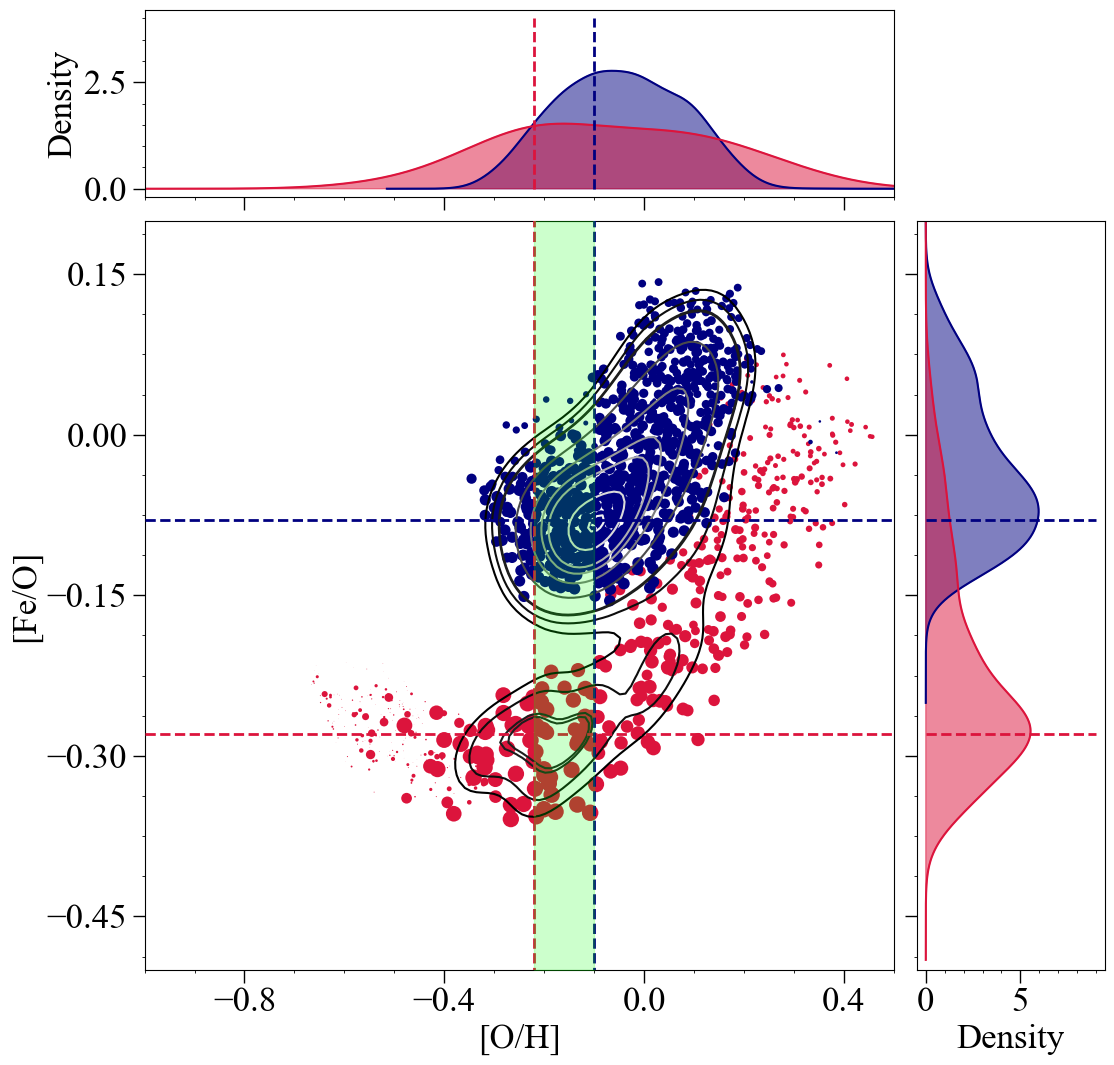}
  \caption{ 'Synthetic model' predictions. Stellar distributions in the [$\alpha$/Fe] versus [Fe/H] (left panels) and [Fe/$\alpha$] versus [$\alpha$/H] (right panels) spaces for $\alpha$ = Mg (upper panels), $\alpha$ = Si (middle panels) and  $\alpha$ = O (lower panels) predicted by the two-infall model taking into account the average observational errors on abundance ratios (see Section \ref{model_stop})  for the high-$\alpha$ phases (red points and distributions, spanning evolutionary time $t<$ T$_{max}$) and for the low-$\alpha$ ones (blue lines and distributions, for  $t\ge$ T$_{max}$). The point sizes are proportional to the number of stars formed at each Galactic time $t$. Moreover, each distribution is normalised in order that its area is 1. In the right panels, the shaded light-green area spans the region of the quantity $\Delta$[$\alpha$/H]$_{\rm peak}$
as defined in eq. (\ref{eq:peak}).}
\label{mod_2phases}
\end{figure*}

\begin{figure*}
\centering
 \includegraphics[scale=0.24]{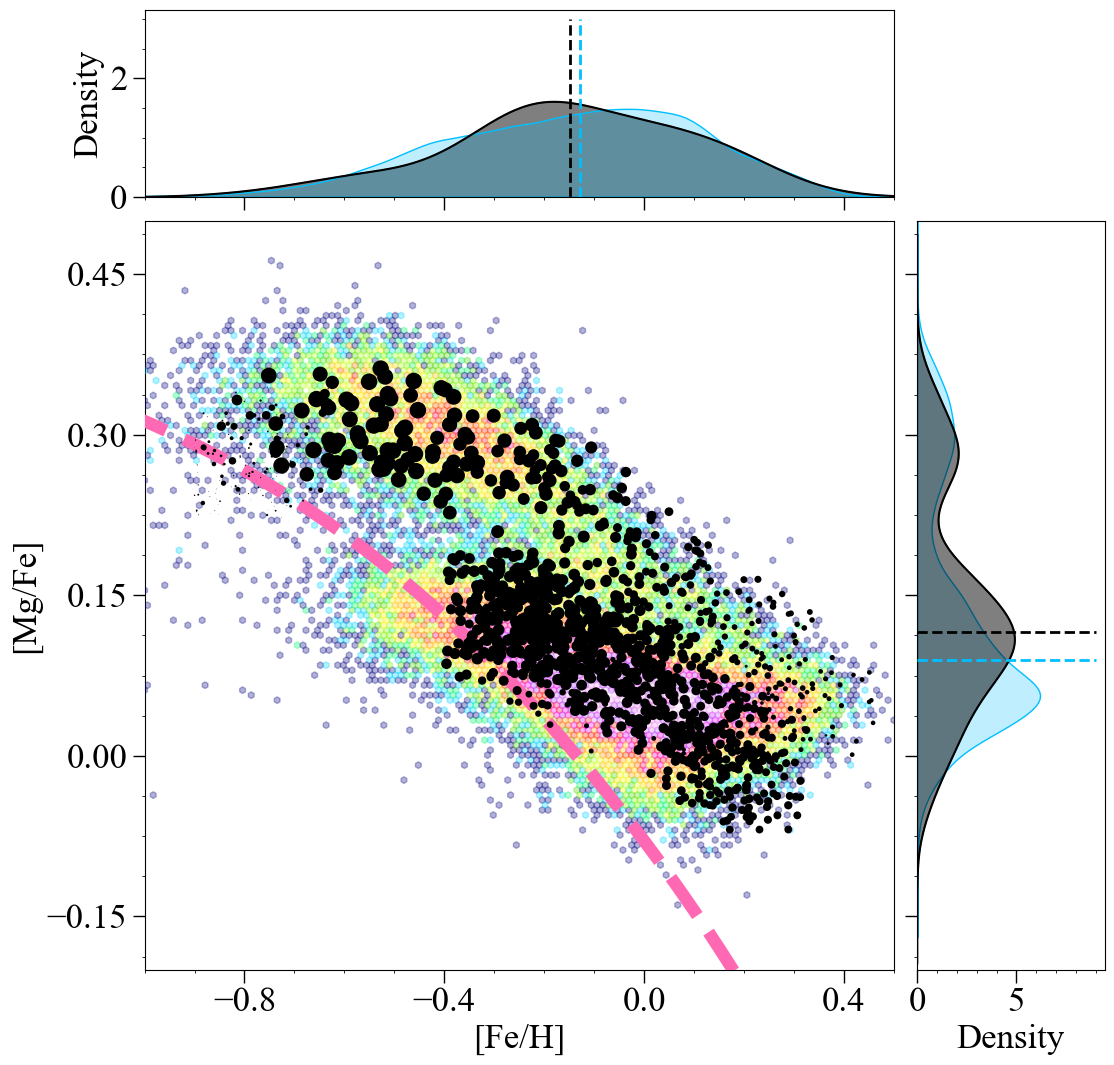}
 \includegraphics[scale=0.24]{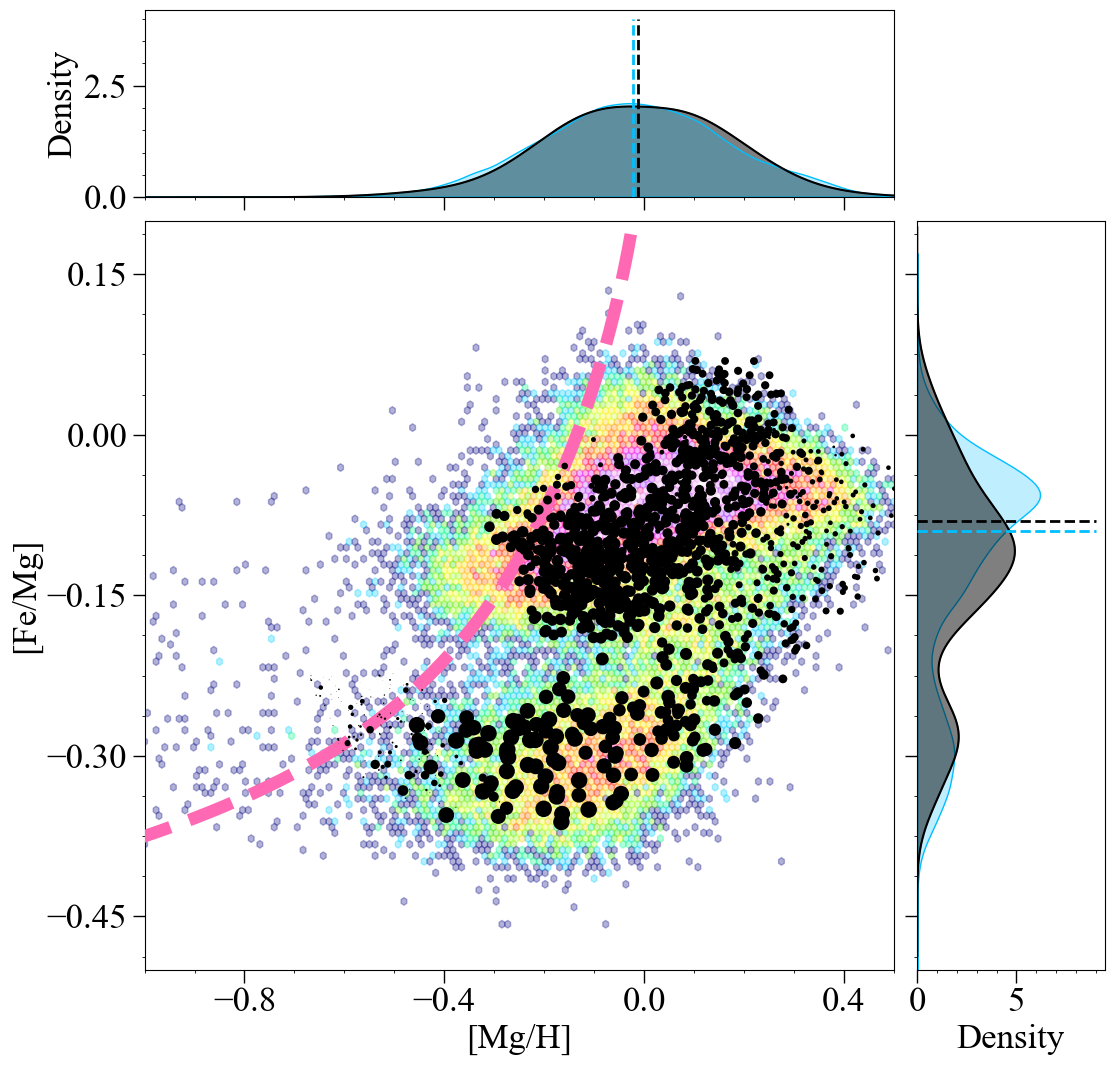}
  \includegraphics[scale=0.24]{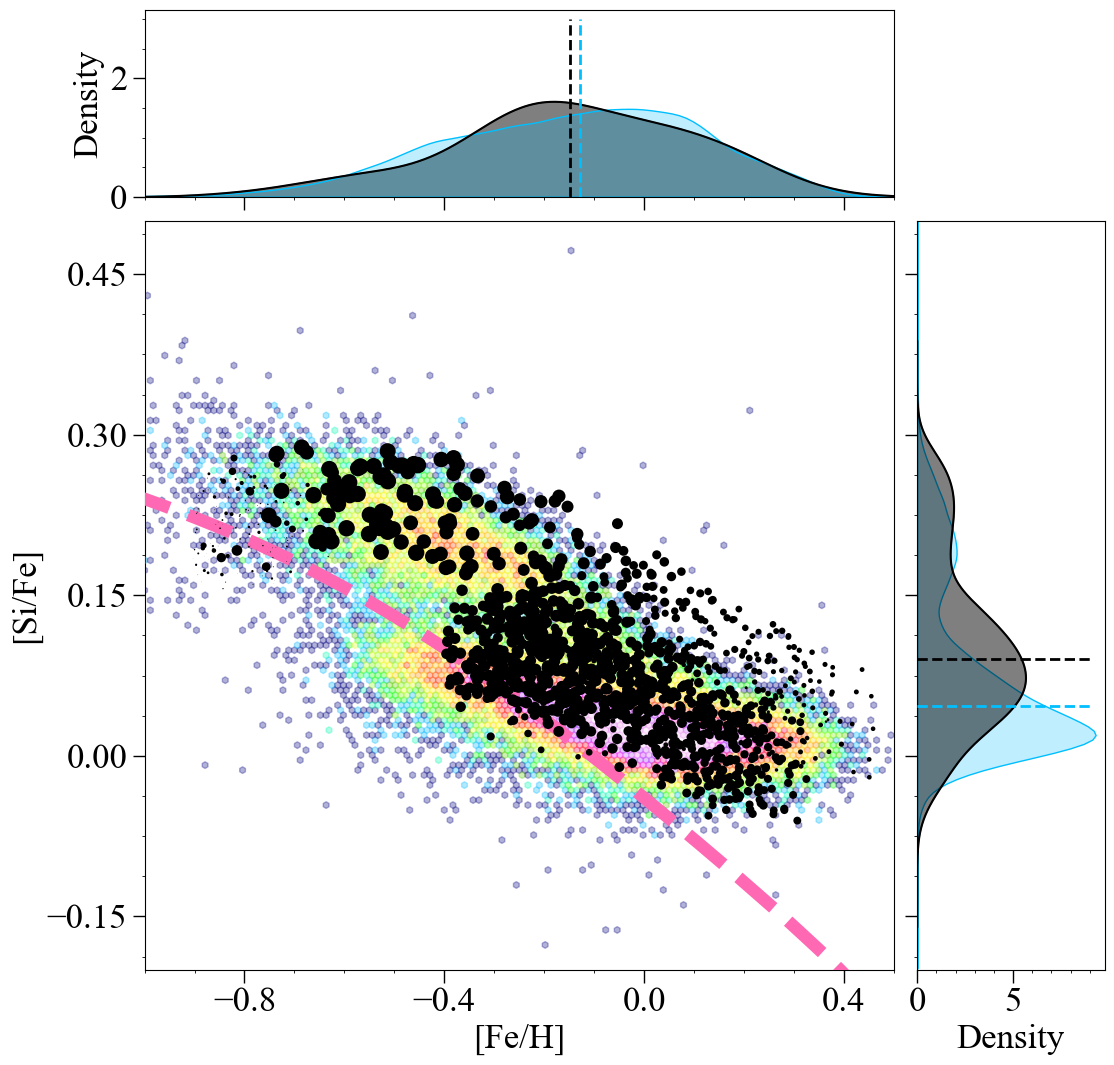}
   \includegraphics[scale=0.24]{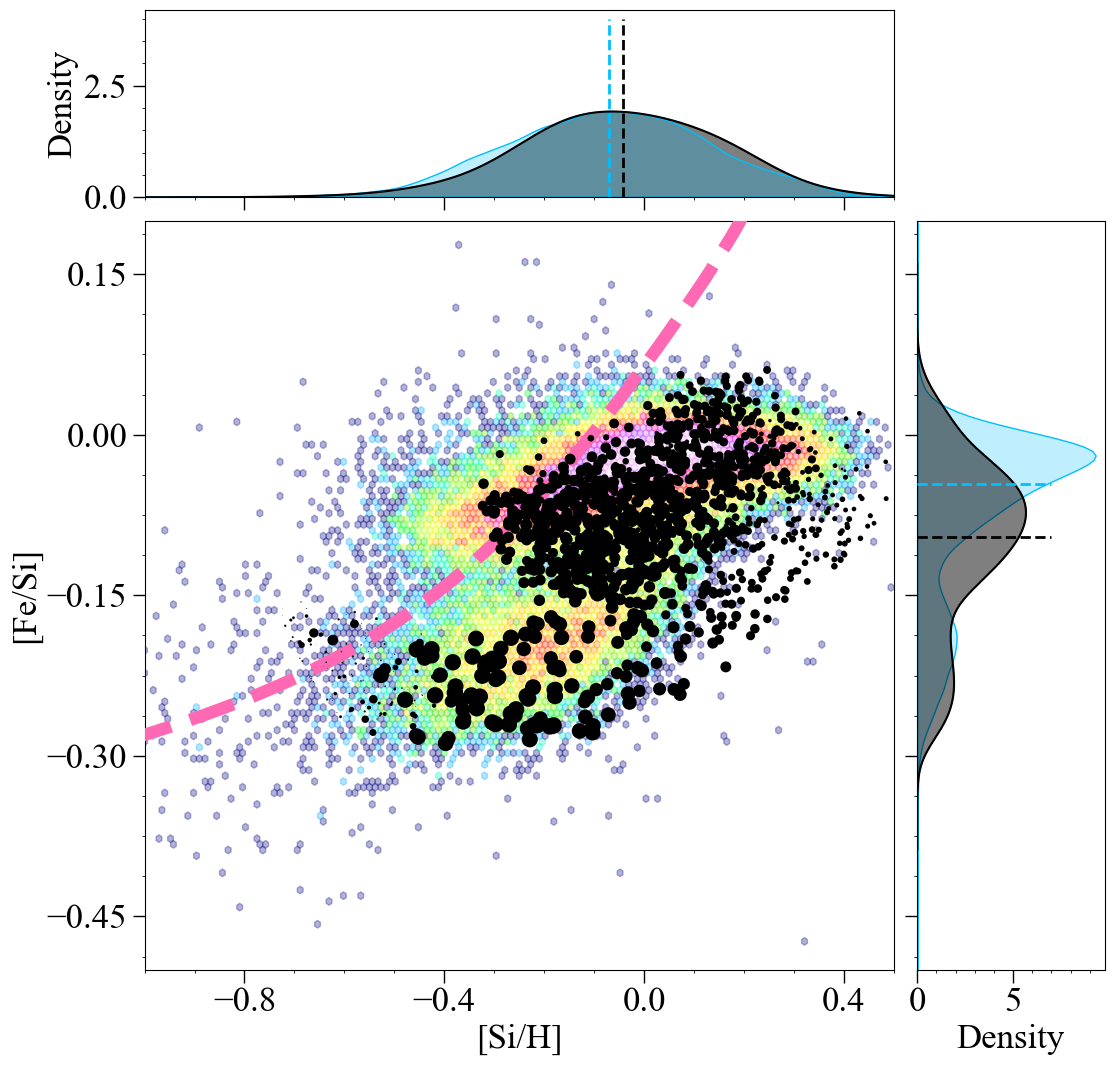}
 \includegraphics[scale=0.24]{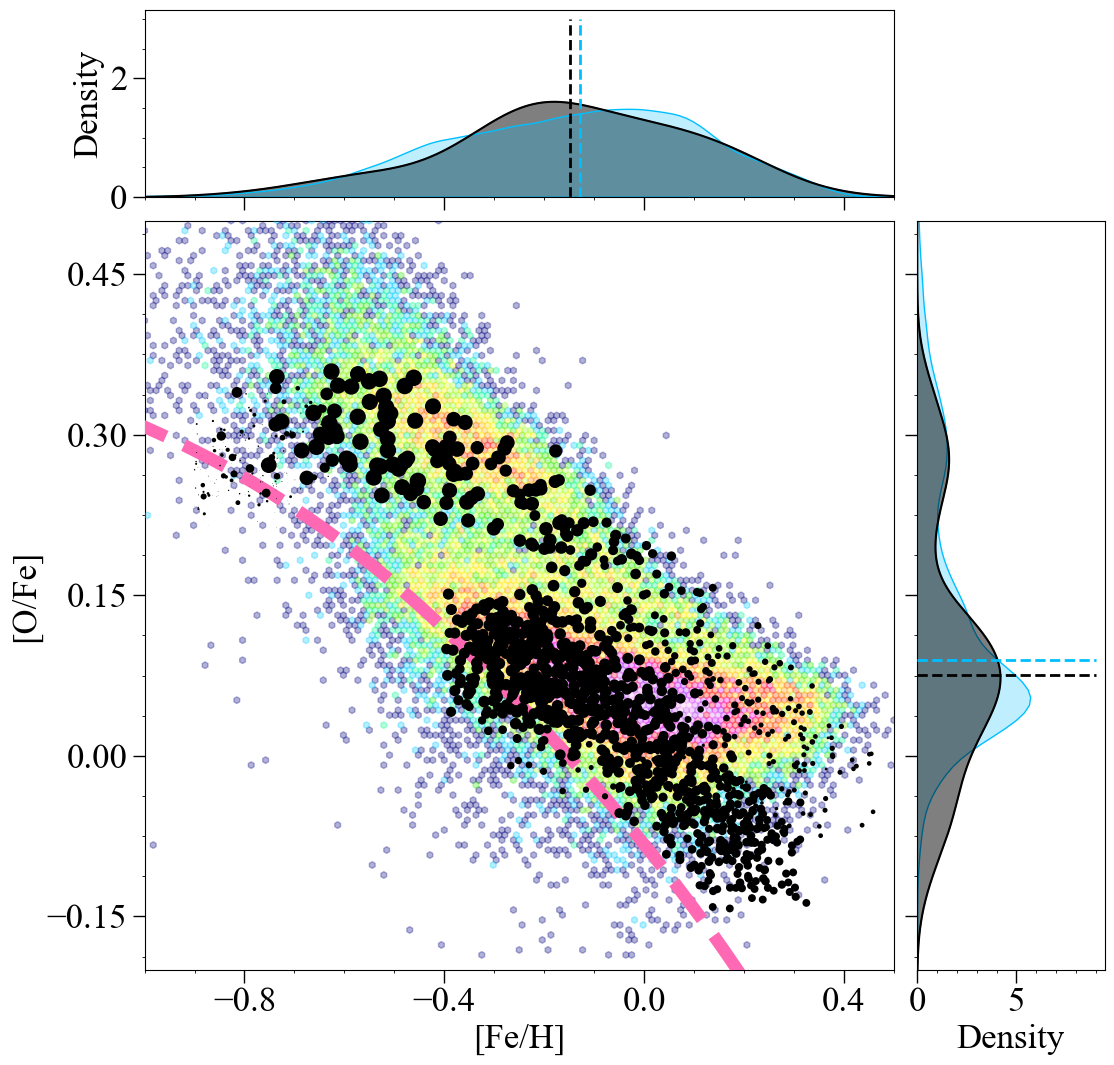}
 \includegraphics[scale=0.24]{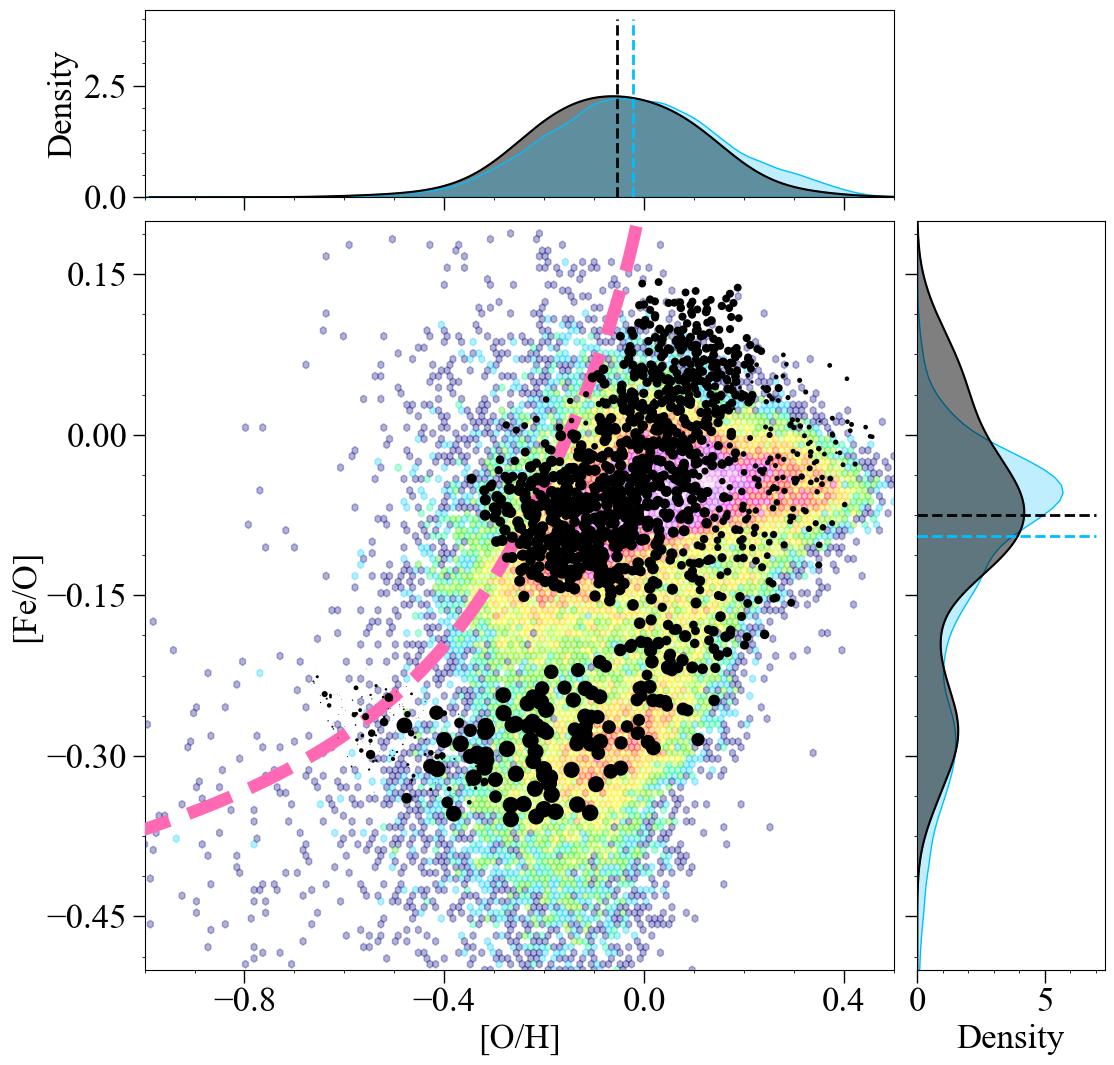}
  \caption{ As in Fig. \ref{fig_Mg_Si_O_model}, but  model predictions  are  indicated with black points   and are referred to  the  'synthetic' model introduced in Section \ref{model_stop}. The point sizes are proportional to the number of stars formed stellar  at different Galactic times.   Model predictions of the massive dwarf galaxy (see Section \ref{dwarf_sec} for model details) are reported with  pink dashed  lines. }
\label{black}
\end{figure*}

For example, in \citet{spitoni2019} the late accretion of pristine gas  produced a dilution of the chemical abundances of the metals thus producing an almost constant [$\alpha$/Fe] ratio and a decreasing [Fe/H],  leading to a horizontal evolution of that ratio.
On the other hand, in the current model, the accretion of chemically enriched gas has the effect of  slightly increasing the slope during the aforementioned horizontal dilution phase.
When star formation resumes, the CC-SNe induce a sharp increase of the [$\alpha$/Fe] ratio, subsequently followed by a decrease and a shift towards higher metallicities due to the strong contamination  in Fe from Type Ia SNe. This sequential pattern results in a distinctive loop in the [$\alpha$/Fe] versus [Fe/H] plane within the chemical evolution track, exhibiting a ribbon-like shape.

In conclusion, we note that for the analysed elements the bimodality in the [$\alpha$/Fe] ratios  and the MDFs for the different elements are well reproduced.
However,  the peak of the distribution of the [$\alpha$/Fe] ratios in the low-$\alpha$ sequences  is slightly shifted towards higher values compared to APOGEE data.  In Section \ref{model_stop}, we will show that the inclusion of observational errors in the predicted abundance ratios improves the  distribution of the [$\alpha$/Fe] ratios for low-$\alpha$ stars, especially for the Mg. 
In this study, we chose not to conduct a detailed analysis of the model parameter space for the delayed infall model scenario, as this has been already extensively discussed in \citet{spitoni2019}. Instead, we retain most of the model parameter values from the study by \citet{nissen2020}, in which we employed a Bayesian approach based on MCMC methods. Notably, these  values of the parameters are also consistent with those used in \citet{spitoni2021} where APOGEE DR16 was used as data constraint again in a Bayesian approach.

\begin{figure*}
\centering
\includegraphics[scale=0.42]{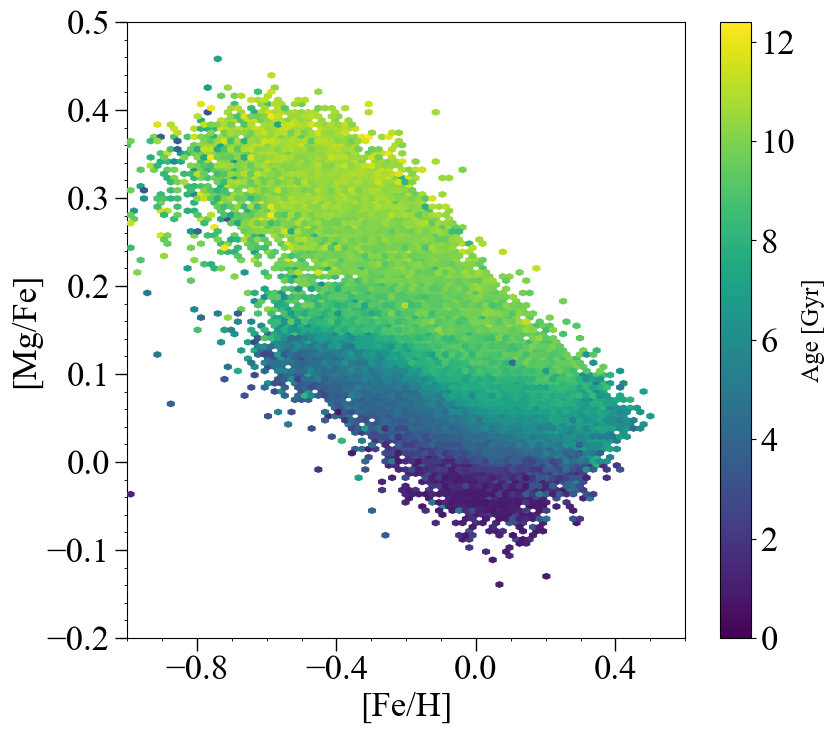}
 \includegraphics[scale=0.42]{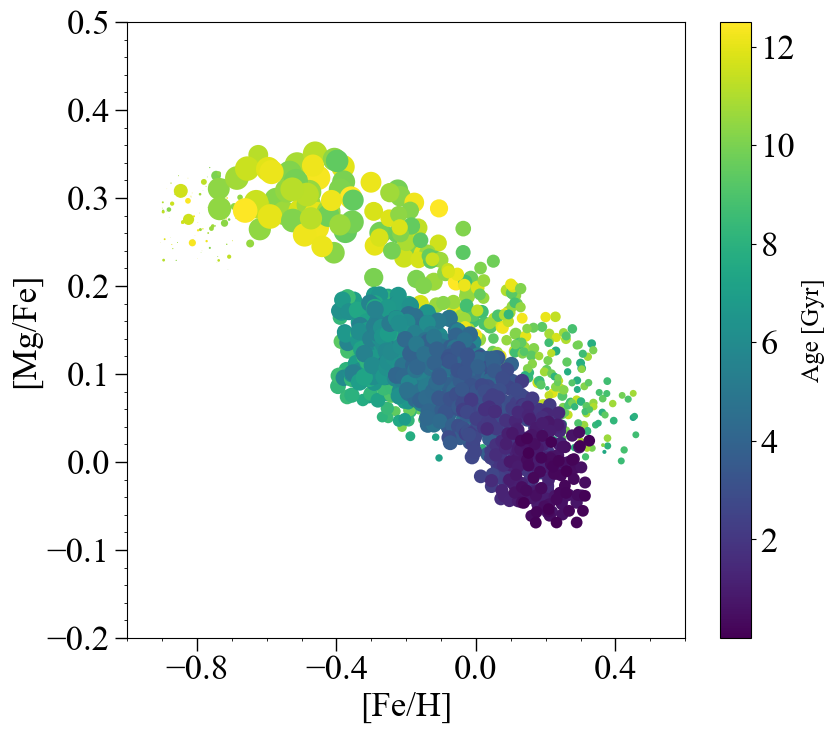} 
\includegraphics[scale=0.42]{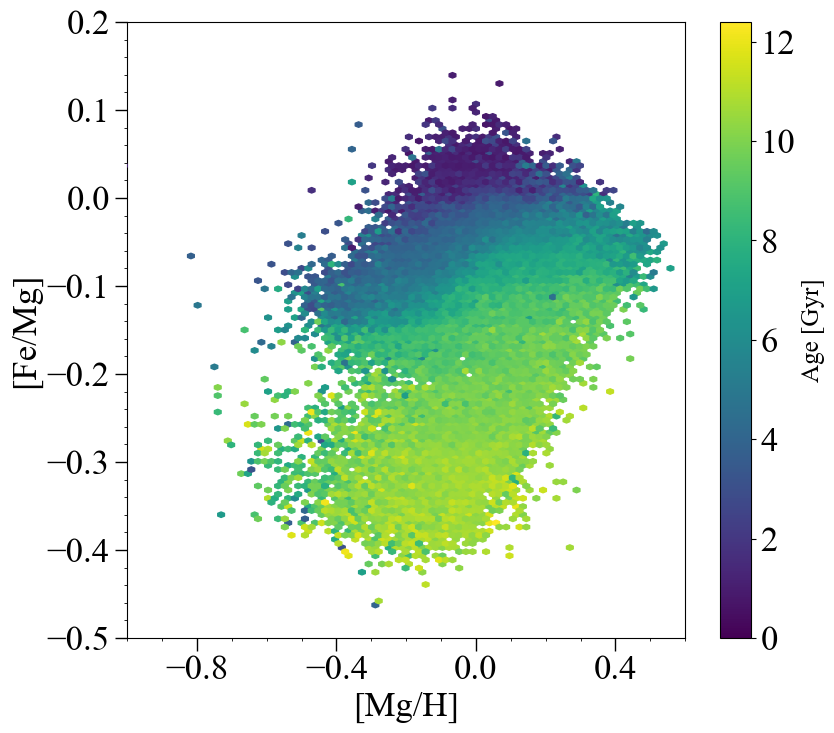}
 \includegraphics[scale=0.42]{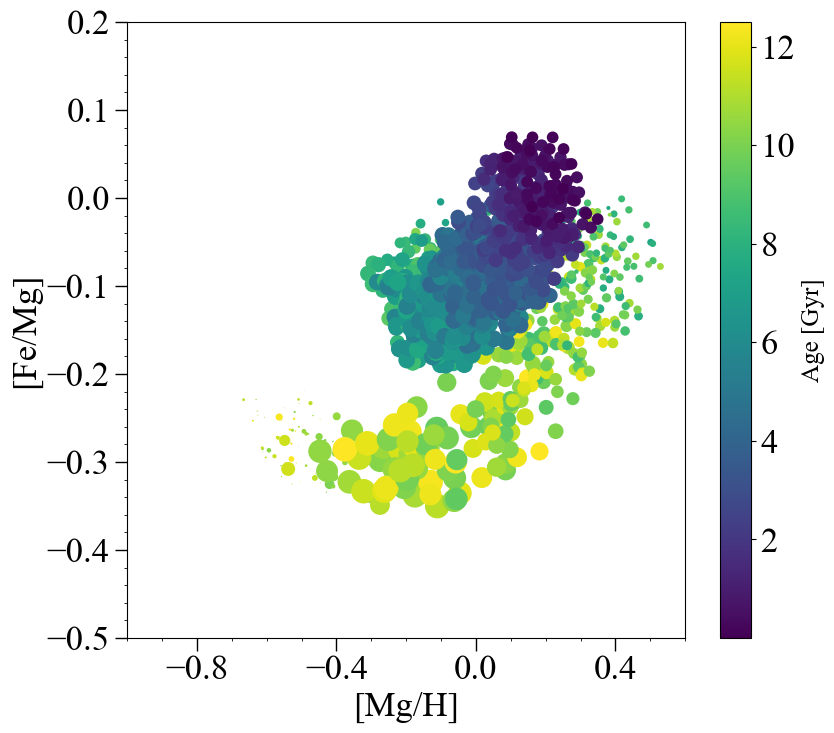}  \caption{ 'Synthetic model' predictions compared to \citet{anders2023} stellar ages.  Upper panels: In the left panel, APOGEE DR17 stars in the [Mg/Fe] versus [Fe/H] space selected as in Fig. \ref{data_density} are color-coded with the stellar ages computed by \citet[][see Section \ref{ages} for details]{anders2023}. 'Synthetic' model predictions including also  age errors (see Section \ref{ages}) are reported on the right panel. 
 Lower panels: As the upper panel but for the  [Fe/Mg] versus [Mg/H] ratios.
 Concerning model predictions, point sizes are proportional to the
number of stars   formed at different evolutionary times  and points are colour-coded with the predicted stellar age (only stars with ages < 12.5 Gyr are reported). }
\label{ages_fig}
\end{figure*}

\begin{figure*}
\centering
 \includegraphics[scale=0.35]{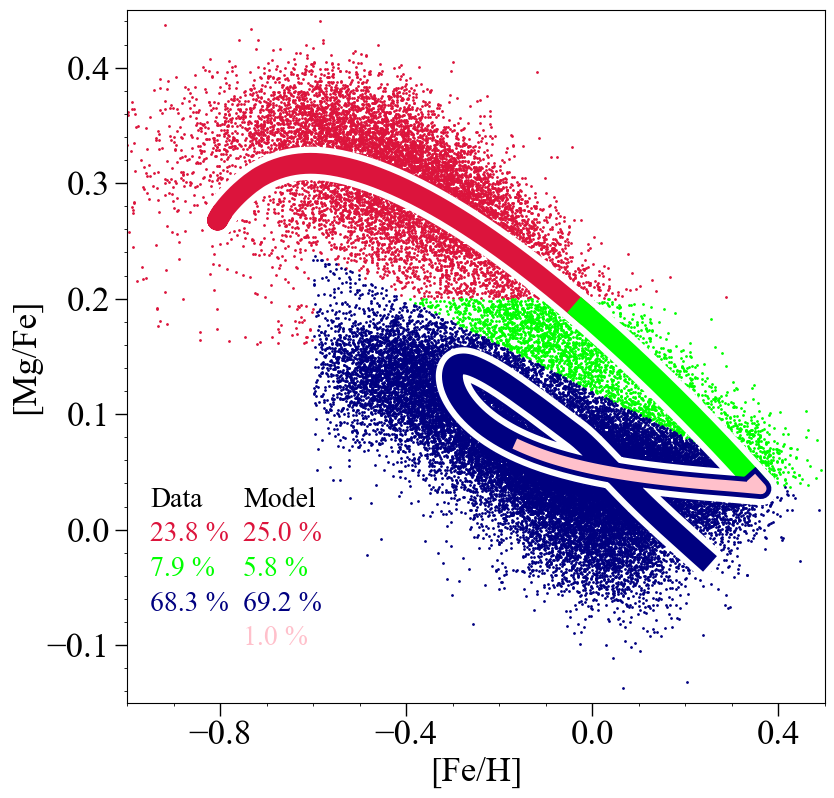}
 \includegraphics[scale=0.35]{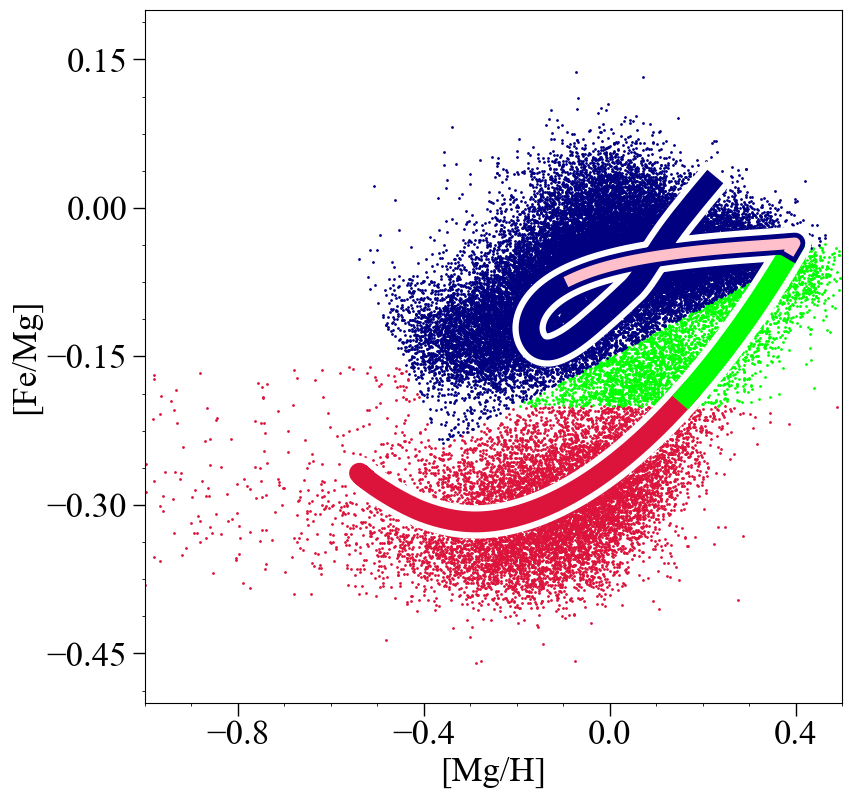}
  \caption{  Counting of the stars throughout the high-$\alpha$ and low-$\alpha$
sequences. Comparison between predicted and observed distributions of stars in 3 regions of the  [Mg/Fe] versus [Fe/H] (left panel) and [Fe/Mg] versus [Mg/H] (right panel) abundance ratios highlighted in red, green and blue. The respective percentages of stars are reported in the legend in the leftmost plot. For the model prediction, we also indicate the percentage of stars which lie in the dilution phase of the low-$\alpha$ in the chemical evolution tracks highlighted with pink lines.     }
\label{3zones}
\end{figure*}
\begin{figure*}
\centering
 \includegraphics[scale=0.55]{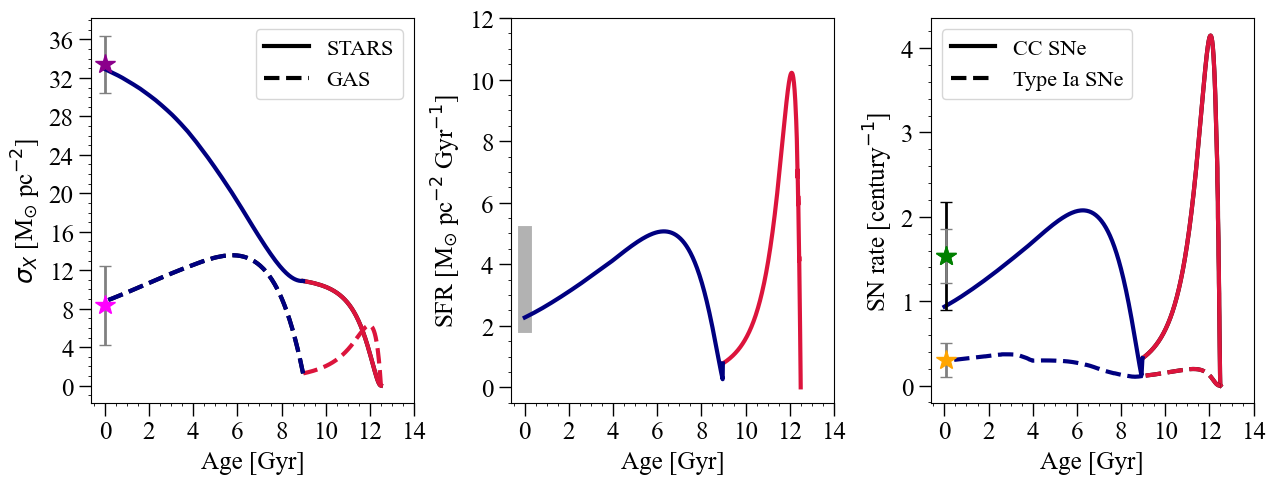}
   \caption{Comparison with Galactic disc present-day observables.
    In each panel, model predictions are drawn with  solid lines, the red and blue colours indicate the high-$\alpha$ (low-$\alpha$) phases, respectively.
   Left panel: Evolution of the surface mass density of stars ($\sigma_{\star}$, solid line) and gas ($\sigma_{\text{gas}}$, dashed line) predicted by the two-infall chemical evolution model presented in this study with model parameters reported in Table \ref{tab_A}. The purple star indicates the observed present-day $\sigma_{\star}$ value given by \citet{mckee2015}. The magenta star represents the present-day $\sigma_{\text{gas}}$ value averaging between the \citet{dame1993} and \citet{Nakanishi2003,Nakanishi2006} datasets as presented by \citet{palla2020} (here shown only at 8 kpc). The 1$\sigma$ errors are reported with grey.  Middle panel: The predicted time evolution of the SFR is shown. The grey shaded area indicates the measured range in the solar annulus suggested by \citet{prantzos2018}.  Right panel: Evolution of the Type Ia SN rate (dashed line) and CC-SN rate (solid line) predicted by the model for the whole Galactic disc. The orange star represents the observed Type Ia SN rate reported by \citet{cappellaro1997}, while the green one represents the observed CC-SN rates from \citet{li2010}. The 1$\sigma$ and 2$\sigma$ errors are reported with grey and black bars, respectively.}
 \label{SN}
\end{figure*}
\subsection{Signatures  of the SF hiatus in the predicted abundance ratios by the 'synthetic' model }
\label{model_stop}

In Fig. \ref{mod_2phases}, we show the predicted distributions in [$\alpha$/Fe] versus [Fe/H] and [Fe/$\alpha$] versus [$\alpha$/H] spaces separating the high-$\alpha$  (evolutionary time $t<$ T$_{max}$) and low-$\alpha$  (evolutionary time $t>$ T$_{max}$) phases,  to analyse in more detail the signature of the star formation hiatus. 
 To be consistent with the definition of $\Delta$[$\alpha$/H]$_{\rm peak}$ introduced in Section \ref{peaks_obs}  for APOGEE DR17 abundance ratios,   we  created a 'synthetic' model considering observed chemical abundance errors  following the same methodology presented in  Section 5 of \citet{spitoni2019}. We  introduced  a random error to the chemical abundances [X/H] (where X=Mg, Si, O, and Fe) of stellar populations formed at each Galactic time $t$. These random errors are uniformly distributed within the range defined by the average errors $\sigma_{[X/H]}=0.1$ dex \citep{spitoni2019} and the  the new predicted abundances can be expressed as follows:

\begin{equation} 
 [\mbox{X/H}]_{new}(t) =  [\mbox{X/H}](t)+  U([-\sigma_{[X/H]}, \sigma_{[X/H]}
 ]).
\label{}
\end{equation}
The same procedure has been applied to [X/Fe] abundance ratios imposing $\sigma_{[X/Fe]}=0.05$ dex \citep{spitoni2020}.
In Fig. \ref{mod_2phases}, it is evident that the  predicted stellar distribution in the [Fe/$\alpha$] versus [$\alpha$/H] plane by the 'synthetic' model is in nice agreement with the APOGEE DR17 data showing the characteristic 'mushroom' shaped stellar distribution of Fig. \ref{data_density}.

From left panels of Fig. \ref{mod_2phases},  we find that $\Delta$[Fe/H]$_{\rm peak} \sim 0.33$ dex.
Examining the right panels for the studied $\alpha$-elements, we observe the following variations: $\Delta$[Mg/H]$_{\rm peak}$ = 0.15 dex, $\Delta$[Si/H]$_{\rm peak}$ = 0.25 dex, and $\Delta$[O/H]$_{\rm peak}$ = 0.11 dex. 
Hence, we have that $\Delta$[$\alpha$/H]$_{\rm peak}$ <<
$\Delta$[Fe/H]$_{\rm peak}$. Moreover, the smallest variation occurs for oxygen, while silicon exhibits the largest $\Delta$[$\alpha$/H] value.  This fact is explained by the fact that Si is produced in a non-negligible way by Type Ia SNe, as already pointed out. In conclusion, these values are in very good agreement with the findings from APOGEE DR17 data presented in Section \ref{peaks_obs}.

Finally, in Fig. \ref{black} we show the effect of the inclusion of observational errors on the abundance ratio distributions without distinguishing between high- and low-$\alpha$ sequences  similarly to what presented in Figure 8 of  \citet{spitoni2019}. We note that from the upper panels, the agreement between [Mg/Fe] distributions between model predictions and APOGEE DR17 has improved compared to Fig. \ref{fig_Mg_Si_O_model}.

\subsection{Comparison with \citet{anders2023} stellar ages}
\label{ages}
 In this Section, we also compare our  model predictions with the recent stellar ages presented by \citet{anders2023}.   Applying a supervised machine learning technique  trained on a high-quality dataset of 3060 red-giant and red-clump stars with asteroseismic ages observed by both APOGEE and Kepler, they  estimated spectroscopic stellar ages for 178825 red-giant. 

In particular, in our analysis, we considered the calibrated spectroscopic age reported in their Figure 4. In the left panel of Fig. \ref{ages_fig},
APOGEE DR17 stars as  selected in Section \ref{apogee}, are color-coded with the median values of the above-mentioned stellar ages.
To properly compare  the  'synthetic model'  introduced in the previous Section  with APOGEE DR17  stellar ages, we included also age errors in our model.   Following the same methodology introduced in  Section \ref{model_stop},   we have that:

\begin{equation} 
 [\mbox{Age}]_{new}(t) =  [\mbox{Age}](t)+  U([-\sigma_{\mbox{Age}}, \sigma_{\mbox{Age}}
 ]),
\label{}
\end{equation}
where for the  $\sigma_{\mbox{Age}}$ quantity we consider 20\%,  value  consistent with the median statistical uncertainty of 17\%
claimed by \citet{anders2023}.

From Fig. \ref{ages_fig}, it is possible to appreciate that our model is capable of capturing the main features of the stellar age distributions both  in the  [Mg/Fe] versus [Fe/H] and  [Fe/Mg] versus [Mg/H] spaces. For clarity, after accounting for age errors, ages greater than 12.5 Gyr were assigned to a small number of predicted stars. Excluding these few objects, the range of ages predicted by our model closely matches the observed range.

In Figure 15 of \citet{chen2023}, the authors presented their model's age distribution in the [Mg/Fe] versus [Fe/H] space. Their results showed that the youngest stars are uniformly distributed across a wide range of metallicities [Fe/H], approximately between -0.5 to 0.4 dex. However, in the APOGEE DR17 data, as shown in Fig. \ref{ages_fig}, the youngest objects are more concentrated in a narrow region between 0 and 0.3 dex,  in agreement with our findings.  

\subsection{Counting of the stars throughout the high- and low-$\alpha$ sequences}
\label{count_stars}

In Fig. \ref{3zones}, we separate the APOGEE DR17  stars in the solar neighbourhood into three zones in the [Mg/Fe] versus [Fe/H] plane.
Stars most likely belonging to the hiatus region are highlighted in green, while the bulk of the high- and low-$\alpha$ sequence stars are denoted in red and blue, respectively. Notably, our chemical evolution model reproduces remarkably well this chemical dissection of the disc. Specifically, the percentages of stars in the red, green, and blue zones are 23.8\%, 7.9\%, and 68.3\% for the data, and 25.0\%, 5.8\%, and 69.2\% for the models.

Moreover, in Fig. \ref{3zones}, we emphasise the percentages of stars predicted in the dilution phase of the low-$\alpha$. It is worth noting that the "almost" horizontal phase in the [Mg/Fe] versus [Fe/H] relation is characterised by a negligible number of stars (i.e., $\sim$ 1\% of the total),  because of the effect of dilution by gas infall occurs mainly in the period of strongly depressed star formation. Therefore,  we do not reproduce the thin disc data, only because of diluted stars, as claimed by \citet{agertz2021}.
Actually, our model results are similar to those of the VINTERGATAN simulations  of \citet{agertz2021} and do not differ from those of \citet{chiappini1997} except for the length of the star formation gap ($\sim$ 1 Gyr in \citealt{chiappini1997} model).  
Hence, the proposed model would be still valid even if the bulk of the high-$\alpha$ stars ( and some of the low-$\alpha$ sequence stars) would be migrators \citep{sharma2021,ratcliffe2023} from the innermost disc. In \citet{spitoni2021}, we have shown that the high-$\alpha$ sequence is similar at different Galactocentric distances, due to the fast formation of the thick disc.

\begin{table}
\begin{center}
\tiny
\caption{Theoretical and observed solar abundances. }
\label{tab_solara}
\begin{tabular}{c|cc}
  \hline
\\
 Element abundance &Data &   Model MW\\
 $\log$($X$/H)+12 &\citet{grevesse2007}&\\
 & [dex]&[dex]\\

\hline
\\
Fe &7.45$\pm$0.05&7.37 \\
 \\
Si &7.51$\pm$0.04& 7.50  \\
 \\
Mg& 7.53$\pm$0.09 & 7.54  \\
 \\
O& 8.66$\pm$0.05 &8.63   \\
 \hline
\end{tabular}
\end{center}
\end{table}

\begin{figure}
\centering
\includegraphics[scale=0.5]{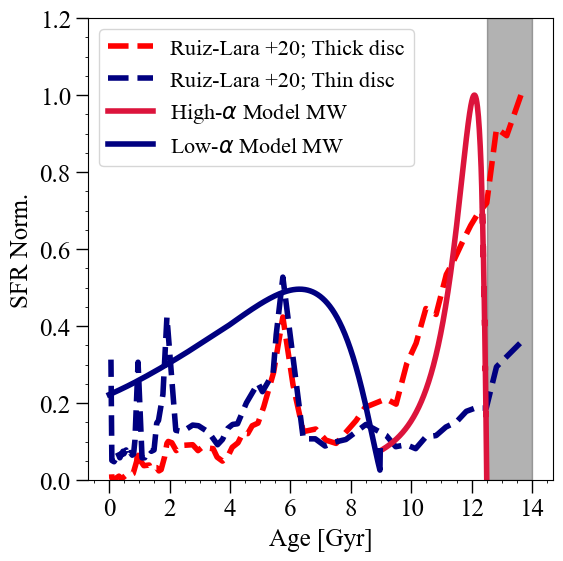}
  \caption{ Temporal evolution of the SFR normalised to the maximum value predicted by our model is drawn with the solid line, the red and blue colours indicate the high-$\alpha$ (low-$\alpha$) phases, respectively  based on chemical selection. The observed estimate obtained with Gaia DR2  by \citet{lara2020}, for the thick (thin) disc  stars  based on kinematic selection  is represented by the red (blue) dashed line. The grey dashed area indicates the Galactic ages that fall outside the evolutionary times of the disc components considered in this study.  }
 \label{lara}
 \end{figure}
 
\subsection{Comparison with Other Galactic disc observables}
\label{other_obs}
In Table \ref{tab_solara}, the solar abundances of Fe, Mg, Si and O as predicted by our two-infall model are compared to \citet{grevesse2007} solar values. The model solar abundances are derived from the composition of the ISM at the time of the formation of the Sun (4.6 Gyr ago). Noteworthy, our model reproduces the solar abundances for the elements considered in this work.
As outlined in Section \ref{2infall_sec}, we impose the condition that the present-day (at evolutionary time $t_g$=13.8 Gyr)  total surface mass density in the local disc, including contributions from both high- and low- $\alpha$ sequences, agrees with the value proposed by \citet{mckee2015} specifically, 47.1 $\pm$ 3.4 M$_{\odot} \mbox{ pc}^{-2}$. 
This study also provides the present-day total local surface density of stars, denoted as $\sigma_{\star}(t_g)$, which is reported as 33.4 $\pm$ 3 M$_{\odot} \mbox{ pc}^{-2}$.

The left panel of Fig. \ref{SN} illustrates the temporal evolution of the surface mass density of stars as predicted by our chemical evolution model. Our model yields $\sigma_{\star}(t_g) = 32.9$ M$_{\odot} \mbox{ pc}^{-2}$, closely matching the value proposed by \citet{mckee2015}. Additionally, the panel displays the predicted temporal evolution of the surface gas density, with the present-day value of $\sigma_g(t_g)$ calculated as 8.77 M$_{\odot} \mbox{ pc}^{-2}$, confirming the value proposed by \citet{palla2020} of 8.4$^{+4.0}_{-4.2}$ M$_{\odot} \mbox{ pc}^{-2}$ obtained by  averaging between the \citet{dame1993} and   \citet{Nakanishi2003,Nakanishi2006} data sets at 8 kpc.The middle panel of Fig. \ref{SN} illustrates the time evolution of the SFR in our model. The present-day value of 2.27 M$_{\odot}$ pc$^{-2}$ Gyr$^{-1}$ lies in the range of 2-5 M$_{\odot}$ pc$^{-2}$ Gyr$^{-1}$ used as a constraint in chemical evolution models for the solar vicinity \citep{matteucci2012,prantzos2018}.
In the right panel of Fig. \ref{SN}, we present the time evolution of Type Ia and CC-SN rates. The proposed model for the Galactic components predicts a present-day CC-SN rate for the entire Galactic disc as 0.93 /[100 yr],  lower (but within a 2$\sigma$ error) than the observations of \citet{li2010} (1.54 $\pm$0.32 /[100 yr]). The predicted present-day Type Ia SN rate for the entire Galactic disc is 0.29 /[100 yr], in perfect agreement with the value provided by \citet{cappellaro1997} (0.30$\pm$0.20 /[100 yr]). Furthermore, the calculated present-day infalling gas rate is 0.68 M$_\odot$ pc$^{-2}$ Gyr$^{-1}$, consistent with the 0.3-1.5 M$_{\odot}$ pc$^{-2}$ Gyr$^{-1}$ range suggested in \citet{matteucci2012} for the solar vicinity.

Finally, in Fig.  \ref{lara} we compare the predicted star formation  history by the two-infall model with the one reconstructed by \citet{lara2020} using Gaia DR2 data (for the thin and thick disc stars), and we note a general agreement with our proposed star formation history.   In \citet{lara2020} the observed peaks in the star formation are attributed to encounters with the Sagittarius galaxy.
It is worth mentioning that to reproduce the observed recent peaks of SF  in the thin disc, \citet{spitoni2023} assumed two  infall episodes occurring in the thin disc. However, the most recent accretion event  affects  only the chemical properties of stars younger than 2.7 Gyr and  does not alter the conclusions of this paper. 
 Finally, it is worth noting that the predicted SFR aligns well with the results obtained by \citet{gallart2024} using the colour–magnitude diagram fitting technique. Both our model and their findings indicate a peak in star formation approximately 12 billion years ago, as illustrated in Fig. \ref{lara}.

\section{Conclusions}
\label{conclu_sec}

In this study,  we  "(re)minded the gap"  in star formation as  proposed by \citet{gratton1996,gratton2000} and \citet{fuhr1998}, based on old stellar samples, specifically focusing on the [Fe/$\alpha$] versus [$\alpha$/H] relation, where $\alpha$ = Mg, Si, O. Our aim was to investigate if signatures of hiatus in the star formation rate are imprinted in the recent spectroscopic data of APOGEE DR17. We  compared data and predictions as derived from a modified version of the two-infall chemical evolution model  by \citet{spitoni2019}, which takes into account the possible merger of the Milky Way with a massive dwarf galaxy, such as Enceladus,  that occurred at early times.

Our key results can be summarised as follows:

\begin{itemize}

\item The APOGEE DR17  [Fe/$\alpha$] versus [$\alpha$/H] relation  exhibits a sharp increase of [Fe/$\alpha$] at nearly constant [$\alpha$/H] during the transition between the two disc phases. This observation supports the hypothesis that a hiatus in star formation occurred during this evolutionary phase. Notably, the most pronounced growth in the [Fe/$\alpha$] versus [$\alpha$/H] relation is observed for oxygen, as this element is exclusively synthesised in CC-SNe. 
In fact, if star formation steadily decreases or stops, oxygen is no more produced whereas iron continues to be ejected by Type Ia SNe acting also in absence of star formation. In this situation [Fe/O] is increasing at constant [O/H].
This confirms the previous suggestions by \citet{gratton1996,gratton2000} and  \citet{fuhr1998}.

\item Our present  two-infall chemical evolution model successfully reproduced the APOGEE DR17 [Fe/$\alpha$] versus [$\alpha$/H] abundance ratio. Particularly noteworthy was the model ability to predict the hiatus in the SFR between the two infalls of gas, generating abundance ratios compatible with APOGEE DR17 data.

\item The present chemical evolution model well reproduces the fractions of APOGEE DR17 stars situated in various regions of the [Mg/Fe] versus [Fe/H] abundance ratio plane. The adopted model predicts two main gas infall episodes forming the thick and thin disc, respectively, separated by a gap lasting $\sim$ 3.5 Gyr. The infalling gas is assumed to be pre-enriched with [Fe/H]=-0.8 dex, in order to account for a merging event with a dwarf galaxy, such as Enceladus,   which occurred   $\sim $1.3 Gyr  after the beginning of evolution.

\item It is important to highlight that the "almost" horizontal evolution in the [Mg/Fe] versus [Fe/H] relation, as predicted by the present two-infall model is characterised by a negligible number of stars, accounting for only 1\% of the total population  since it represents the phase where the star formation is strongly depressed. Therefore, the gas infall dilution does not play an important role on the formation of the majority of low-$\alpha$ stars in our model, at variance with what claimed by \citet{agertz2021} when commenting  the \citet{spitoni2019} model.
\end{itemize}
\section*{Acknowledgement}
 The authors thank the anonymous referee for the valuable suggestions that improved the paper.
E. Spitoni and G. Cescutti thank I.N.A.F. for the  
1.05.23.01.09 Large Grant - Beyond metallicity: Exploiting the full POtential of CHemical elements (EPOCH) (ref. Laura Magrini).
 F. Matteucci thanks I.N.A.F. for the 1.05.12.06.05 Theory Grant - Galactic archaeology with radioactive and stable nuclei. This research was supported by the Munich Institute for Astro-, Particle and BioPhysics (MIAPbP) which is funded by the Deutsche Forschungsgemeinschaft (DFG, German Research Foundation) under Germany´s Excellence Strategy – EXC-2094 – 390783311. F. Matteucci thanks also support from Project PRIN MUR 2022 (code 2022ARWP9C) "Early Formation and Evolution of Bulge and HalO (EFEBHO)" (PI: M. Marconi), funded by the European Union – Next Generation EU. 
B. Ratcliffe  and I. Minchev acknowledge support by the Deutsche Forschungsgemeinschaft under the grant MI 2009/2-1. This work was also partially supported by the European Union (ChETEC-INFRA, project number 101008324). 
 In this work, we have made use of SDSS-IV APOGEE-2 DR17 data. Funding for the Sloan Digital Sky Survey IV has been provided by the Alfred P. Sloan Foundation, the U.S. Department of Energy Office of Science, and the Participating Institutions. SDSS-IV acknowledges
support and resources from the Center for High-Performance Computing at
the University of Utah. The SDSS web site is  \href{www.sdss.org}{www.sdss.org}.
SDSS is managed by the Astrophysical Research Consortium for the Participating Institutions of the SDSS Collaboration which are listed at \href{https://www.sdss.org/collaboration/affiliations/}{www.sdss.org/collaboration/affiliations/}.

\bibliographystyle{aa} 
\bibliography{disk}

\end{document}